\documentclass[preprint,12pt,authoryear]{elsarticle}

\usepackage{amssymb}
\usepackage{amsmath}
\usepackage{amsthm}
\usepackage{bm}
\usepackage{mathrsfs}

\usepackage{algorithmic}
\usepackage{algorithm}
\usepackage{xcolor}
\newtheorem{theorem}{Theorem}
\usepackage{multirow}

\theoremstyle{definition}
\newtheorem{definition}{Definition}

\theoremstyle{remark}
\newtheorem{remark}{Remark}

\newcommand{\clr}{\textrm{clr}}
\newcommand{\ilr}{\textrm{ilr}}
\newcommand{\spansb}{\textrm{span}}
\newcommand{\diag}{\textrm{diag}}
\newcommand{\trace}{\textrm{tr}}

\newcommand{\transf}[1]{\widetilde{\mathbf{#1}}}

\newcommand{\transfDat}[1]{\widetilde{\mathcal{#1}}}

\journal{Computational Statistics \& Data Analysis}

\begin{document}

\begin{frontmatter}



\title{Testing for a common subspace in compositional datasets with structural zeros}

\author[label1]{Francesco Porro}
\author[label2]{Fabio Rapallo}
\author[label1]{Sara\corref{cor1} Sommariva}

\cortext[cor1]{sara.sommariva@unige.it}

\affiliation[label1]{organization={Department of Mathematics, University of Genova},
             addressline={via Dodecaneso 35},
             city={Genova},
             postcode={16146},
             state={},
             country={Italy}}

\affiliation[label2]{organization={Department of Economics, University of Genova},
             addressline={via Francesco Vivaldi 5},
             city={Genova},
             postcode={16126},
             state={},
             country={Italy}}



\begin{abstract}
In this paper, we consider the problem of testing for a common principal subspace in compositional datasets with structural zeros. In particular, we address the problem in the general setting in which two groups are compared, each characterized by its own pattern of structural zeros. This situation prevents the direct use of standard logratio-based principal subspace comparison methods, since the two groups cannot be represented in a common logratio coordinate system in the usual way.

We thus define a test for the presence of a common subspace that is fully compatible with the Aitchison geometry and logratio analysis. The proposed construction allows the principal subspaces associated with the two groups to be compared despite the different patterns of structural zeros. Under logratio-normality, we derive an analytical approximation to the null distribution of the test statistic. Alongside this parametric version, we also introduce a nonparametric bootstrap procedure that does not rely on distributional assumptions. The finite-sample behaviour of the test is investigated through simulations. The method is finally illustrated on a reproducible microbiome dataset available through Bioconductor.
\end{abstract}




\end{frontmatter}



\section{Introduction}
\label{sec1}

The problem we deal with in this paper falls in the framework of dimensionality reduction for compositional data with structural zeros. Compositional data arise in contexts where the relevant information lies in the proportions among the observed variables and not in their values or their sum. Although compositional data analysis was historically developed within specialized application areas, its relevance has substantially increased with the widespread use of microbiome and other high-throughput omics data \citep{tsilimigras2016}. Besides microbiome analysis, other fields of application include tourism \citep{grifoll2019}, finance \citep{fiori2023}, pattern recognition \citep{lu2024}, geochemistry and analytical chemistry \citep{rieser2023, mert2015}.

At the core of compositional data analysis lies the definition of compositions. A composition with $D$ parts is a real-valued vector having all strictly positive components that sum to $1$, i.e., an element of the $(D-1)$-dimensional simplex defined by:
\begin{equation}
\mathbb S^D=\left\{\mathbf{x}=(x_1,\dots,x_D)\;:\;x_i>0, \forall i;\;\sum_{i=1}^D x_i=1 \right\} . \label{eq:simplex}
\end{equation}



A typical compositional dataset $\mathcal{X}$ is a matrix with $n$ rows and $D$ columns that collects a sample of $n$ observations, each one being a $D$-part composition:
\begin{equation}\label{eq:comp_dataset}
\begin{split}
& \qquad\qquad\qquad\mathcal{X} =(\mathbf x_1,\dots,\mathbf x_n)',\;\mbox{where:}\\
&\mathbf x_i=(x_{i1},\dots, x_{iD}) \in \mathbb S^D,\; \sum_{j=1}^Dx_{ij}=1\;\; \; i=1,\dots,n.
\end{split}
\end{equation}

Standard statistical descriptive measures based on the real Euclidean geometry can lead to erroneous conclusions when applied to compositional datasets \citep{pawlowsky2015, mert2015,grifoll2019}. This issue is usually overcome through the so-called \emph{principle of working in coordinates} \citep{mateu2011, grifoll2019}: a proper transformation is defined to map each composition into a vector of coordinates belonging to suitable spaces equipped with a Euclidean structure. Then, standard statistical approaches, such as e.g. Principal Component Analysis (PCA) for dimensionality reduction, can be applied to the transformed data. 

Many transformations based on logratios have been proposed, among which the most common ones are the additive logratio (\rm{alr}), the centered logratio (clr), and the isometric logratio (ilr) \citep{aitchison1982, pawlowsky2015,egozcue2003,mateu2011}.
All these three transformations are based on the ratios of logarithms of parts belonging to a composition. From a theoretical perspective, this does not cause problems since, by definition of composition, and also as shown in the simplex formula in Eq.~\eqref{eq:simplex}, all the parts of a composition must be greater than zero. Unfortunately, in many real-world applications, the presence of a zero part can occur. In those cases, the aforementioned transformations cannot be applied, so new procedures must be considered. 

As detailed in \citet{pawlowsky2015} and \citet{martinfernandez2012}, there are distinct types of zeros. In this paper, we focus on \emph{structural zeros}, which are actual zeros, representing the absence of the phenomenon under analysis. As an example, in a study on monthly expenditure of a set of families, the proportion of expenditure in children school services in families without children is a structural zero. Structural zeros are different from \emph{counting zeros}, which are caused by sampling issues of an unobserved part, or from \emph{rounded zeros}, which are due to a measurement under a certain threshold \citep{pawlowsky2015, kim2024}. For a detailed review on the comparison of zero-replacement strategies for compositional data see \cite{lubbe2021}.

Although it can sound reasonable to replace counting or rounded zeros with a certain (small) value $\epsilon$ \citep{lubbe2021,filzmoser2018}, this procedure does not seem acceptable when dealing with structural zeros. Indeed, in this case a widely used approach consists in splitting the compositional dataset into two or more subsamples, thus isolating groups of observations with the same pattern of structural zeros \citep{pawlowsky2015}. 

Without loss of generality, to illustrate this approach here and in the remainder of the paper we consider a compositional dataset $\mathcal{X}$ divided into two parts, denoted by $\mathcal{Y}$ and $\mathcal{Z}$, which share a common group of non-zero parts and have structural zeros in different columns. Although the precise notation and the meaning of each letter will be introduced later in Section \ref{sec:background}, the structure of the dataset we consider here looks as follows:
\begin{equation*}
\mathcal X= \left( \begin{array}{ccc}
\mathcal{Y}^{(1)} & \mathcal{Y}^{(2)} & \mathbf{0}_{n_y,Q_1} \\
\mathcal{Z}^{(1)} & \mathbf{0}_{n_z,Q_2} & \mathcal{Z}^{(2)}
\end{array}  \right).
\end{equation*}
When performing dimensionality reduction with PCA, the aforementioned classical approach will result in two independent analyses for the two subsamples $\mathcal{Y}$ and $\mathcal{Z}$. However, this may be limiting because, despite the structural zeros, the two subsamples may come from a unique population, and the distinction in two subpopulations can be misleading. From a statistical (or better geometrical) perspective, this can be represented by the situation where the two subsamples still share a common subspace that would allow for a representation of the whole dataset in the same geometrical space and support classification or stratification studies involving the whole population. 

A widespread approach for identifying a principal subspace shared by different groups is the Common Principal Components (CPC) model proposed by \cite{flury1984}, which assumes that the covariance matrices of all groups share the same principal directions while allowing group-specific eigenvalues. Other approaches compare principal component subspaces without requiring complete coincidence of the principal axes \citep{krzanowski1979, schott1988}. However, these methods are designed for datasets lying in the same Euclidean space, hypothesis violated when structural zeros are presented. 

The main contribution of this paper is to address the issue described above. In particular, we generalize the hypothesis test for the existence of a common subspace proposed in \cite{schott1988} to the more complex setting of compositional datasets with structural zeros. Towards this end, first we introduce a suitable test statistic based on the eigenvalues associated to the first, more important principal components of the two subsamples $\mathcal{Y}$ and $\mathcal{Z}$. Then, we analytically approximate the null distribution of the test statistic under the assumption that data are realizations of random compositions normally distributed on the simplex. Furthermore, a  bootstrap procedure is introduced to deal with the non-parametric, distribution-free case. Empirical size and power of the proposed parametric and non-parametric tests are investigated using simulated data, while their effectiveness in real-world scenarios is discussed through their application to an experimental compositional dataset coming from the Human Microbiome Project and available through the \texttt{Bioconductor} package \citep{schiffer}. In detail, we used our approaches to compare the microbiome distribution across different sites proving their capability to identify sites sharing common characteristics in the phyla counts despite presenting different patterns of structural zeros.

The paper is organized as follows. In Section \ref{sec:background} we review the tools for compositional data analysis needed in our work, including the most common logratio transformations, and the classical approach for PCA of compositional data. In Section \ref{sec:proptest} we introduce the proposed statistical test illustrating the computation of the approximated null distribution of the test statistic in the parametric framework and the bootstrap procedure for the non-parametric case. A detailed validation of the proposed approach using simulated data is presented in Section \ref{sec:validationsim}, while in Section \ref{sec:validationreal} we apply our approach to the experimental compositional dataset from the Human Microbiome Project. Finally, our conclusions are offered in Section \ref{sec:discussion}.

\section{Principal component analysis with compositional data}
\label{sec:background}

In this section, we briefly review the basic tools for the analysis and PCA of compositional data. Specifically, in Sections \ref{sub:geometry} and \ref{sub:PCA} we consider the case of compositional data without structural zeros, while in Section \ref{sub:PCAzeros} we introduce the problem of PCA in the presence of structural zeros.

\subsection{Aitchison geometry and logratio transformations} \label{sub:geometry}

From any vector $\mathbf{x} = (x_1,\dots, x_D) \in \mathbb R^D$ with non-negative components, we can obtain a composition by computing the \emph{closure (to 1)} of $\mathbf{x}$, defined as
\[
\mathscr{C}(\mathbf{x})=\mathscr{C}(x_1,\dots,x_D)=\left(\frac{x_1}{\sum_{i=1}^{D} x_i},\dots, \frac{x_D}{\sum_{i=1}^{D} x_i}\right) \in \mathbb S^D.
\]

Moreover, it can be proved \citep{aitchison1982, pawlowsky2015} that $\mathbb S^D$ is a Euclidean $\mathbb{R}$-vector space with the operations of perturbation, powering, and the Aitchison inner product defined, for strictly positive compositions, as 
\begin{equation}
\langle\mathbf x,\mathbf y\rangle_a=\frac{1}{2D}\sum_{i=1}^D\sum_{j=1}^D\left(\ln\frac{x_i}{x_j}\ln\frac{y_i}{y_j}\right) \qquad  \mathbf{x},\mathbf{y}\in \mathbb S^{D}.\label{eq: inner_prod}
\end{equation}


As mentioned in the introduction, the principle of working in coordinates suggests performing a transformation of a compositional dataset. Here are some details on the two most commonly used: the centered logratio (clr) and the isometric logratio (ilr) transformation.

The clr transformation of the $D$-part composition $\mathbf x=(x_1, \ldots ,x_D)$ is defined as
\[
\clr(\mathbf x)=\left(\log \frac{x_1}{g(\mathbf x)}, \ldots ,\log \frac{x_D}{g(\mathbf x)}\right),
\]
where $g(\mathbf x)=\left(\prod_{i=1}^{D} x_i\right)^{1/D}$ denotes the geometric mean of $\mathbf x$. The clr transformation is a function from $\mathbb S^D$ to $\mathbb R^D$, which maps an element of the simplex to a vector whose components sum to 0. Moreover, it preserves distances and angles, meaning that the Aitchison inner product of two compositions in $\mathbb{S}^D$ is equal to the usual Euclidean inner product of the corresponding transformed vectors in $\mathbb R^D$ \citep[for further details, see][and references therein]{pawlowsky2015}.

The ilr transformations are closely related to the orthonormal bases of $\mathbb S^D$ \citep{egozcue2003}: given an Aitchison-orthonormal basis $\{\mathbf{e_1},\ldots ,\mathbf{e_{D-1}}\}$ in the simplex $\mathbb S^D$ the corresponding ilr transformation maps an element $\mathbf x \in \mathbb S^D$ into the vector
\[
	\ilr(\mathbf x)=\bigl(\langle \mathbf{x},\mathbf{e_1}\rangle_a,, \ldots ,\langle \mathbf{x},\mathbf{e_{D-1}}\rangle_a\bigr) \in \mathbb{R}^{D-1} \, .
	\]

From the definition, it clearly follows that several ilr transformations can be considered. Here we focus on the one introduced by \cite{egozcue2003} and related to the orthonormal basis consisting of the vectors  
\[
\mathbf {e_{\textit{\textbf{i}}}}=\mathscr C(\exp(\bm{w_i}))
 \qquad i=1,\ldots ,D-1,
\]
where the vectors $\bm{w_i}$ are given by
\[
\bm w_i=\sqrt{\frac{i}{i+1}}\left( \underbrace{-\frac{1}{i},...,-\frac{1}{i}}_{i},1, 0,...,0\right) \qquad i=1, \ldots ,D-1,
\]
and the exponential function is computed component-wise.
This basis determines the so-called \emph{backward pivot (logratio) coordinates} \citep{nesrstova2023_bpc} of the composition $\mathbf x=(x_1,\dots,x_D)$, denoted in this work by
\[
\transf{x}=\ilr(\mathbf x) =(\tilde{x}_1,\ldots,\tilde{x}_{D-1}),
\]
where each component $\tilde{x}_i$ is 

\begin{equation}\label{eq:def_ilr}
\tilde{x}_i=\sqrt{\frac{i}{i+1}} \log \frac{x_{i+1}}{(\prod_{j=1}^i x_j)^{1/i}},\qquad i=1, \dots ,D-1.
\end{equation}

The backward pivot (logratio) and the clr coordinates are strictly related through the following formulas:
\[
\mathbf{\tilde{x}}=\ilr(\mathbf x)=\mathbf{W}^T\clr(\mathbf x)
\]
\[
\mathbf x=\ilr^{-1}(\mathbf{\tilde{x}})=\mathscr C(\exp(\mathbf{W}\bm{\mathbf{\tilde{x}}})),
\]
being $\mathbf{W}$ a $D\times (D-1)$ matrix whose columns are the vectors $\bm{w}_1, \dots, \bm{w}_{D-1}$.

\subsection{Aitchison’s approach for PCA of compositional data} \label{sub:PCA}

Principal Component Analysis is one of the first statistical methods adapted for compositional data analysis.
The first attempts in that direction are due to \citet{aitchison1983, aitchison1984}, who proposed a logratio-transformation of the data before the application of PCA, basically for two reasons. 
The first one is related to the marked curvature often displayed by a compositional dataset, which cannot be captured by standard principal components. The second one pertains to the constant-sum constraint in compositional data, which imposes structural restrictions on the correlation matrix of the raw proportions. Consequently, the correlations are not entirely unconstrained, and in this context, it seems of little use to insist on Euclidean orthogonality and the zero correlation of linear combinations of raw proportions \citep[see for further details][]{aitchison1983}.

Aitchison's approach has been largely applied in many fields  \citep{cicchella2022,aitchison2002,wang2015}. Effectively, it has revealed itself to be more appropriate in capturing the non-linear curved patterns often displayed by compositional datasets than classical PCA performed on the original data \citep[see for example][]{aitchison1983, aitchison2002, filzmoser2009,filzmoser2018}. In this paper, following the approach suggested by \cite{filzmoser2009}, we apply Aitchison's approach by considering the ilr transformation, since it allows to obtain non-collinear data, and thus full-rank covariance matrices of the data.

A relevant drawback of such a procedure is that it cannot be applied whenever there are data with value equal to zero, because the argument of the logarithm function must be strictly positive. As mentioned in the previous section, this issue may be overcome by replacing the zeros by a (small) positive value and then applying the logratio-transformation to the modified dataset. Nevertheless, in the presence of many zero parts, and especially in the case of structural zeros, the replacement cannot be considered a reasonable choice or a good practice.


\subsection{PCA of compositional data with structural zeros} \label{sub:PCAzeros}
\label{sec3}

This section addresses the issue of how to perform a PCA in the presence of structural zeros in the data. First, we need to introduce the setting that we will use throughout the paper. 

In a compositional dataset, as described in Eq.~\eqref{eq:comp_dataset}, we assume that $n_y$ observations (i.e. compositions) have $Q_1$ parts (with $0<Q_1<D$) that are structural zeros, and the remaining $n_z$ observations have $Q_2$ other parts (with $0 \leq Q_2 < D-Q_1)$ with the same property. Without loss of generality, we can assume:
\begin{equation}\label{eq:comp_data_zeros}
\mathcal X= \left( \begin{array}{ccc}
\mathcal{Y}^{(1)} & \mathcal{Y}^{(2)} & \mathbf{0}_{n_y,Q_1} \\
\mathcal{Z}^{(1)} & \mathbf{0}_{n_z,Q_2} & \mathcal{Z}^{(2)}
\end{array}  \right)
\end{equation}
where
\begin{displaymath}
(\mathcal{Y}^{(1)}, \mathcal{Y}^{(2)}) =: \mathcal{Y} = (\mathbf{y}_1, \dots,\mathbf{y}_{n_y})' \;\; \mbox{and} \;\;  (\mathcal{Z}^{(1)}, \mathcal{Z}^{(2)}) =: \mathcal{Z} = (\mathbf{z}_1, \dots,\mathbf{z}_{n_z})'
\end{displaymath}
being
\begin{displaymath}
\begin{split}
&\mathbf y_i=(y_{i1}, \dots, y_{i(D-Q_1)}) \in \mathbb S^{(D-Q_1)},\;\; \sum_{j=1}^{D-Q_1} y_{ij}=1\;\; \; i=1,\dots,n_y \\
&\mathbf z_i=(z_{i1}, \dots, z_{i(D-Q_2)}) \in \mathbb S^{(D-Q_2)},\;\; \sum_{j=1}^{D-Q_2} z_{ij}=1\;\; \; i=1,\dots,n_z .
\end{split}
\end{displaymath}
Moreover, in this setting, $n_y + n_z = n$, $\mathbf{0}_{n_y,Q_1}$ and $\mathbf{0}_{n_z,Q_2}$ are matrices of size $n_y \times Q_1$ and $n_z \times Q_2$ whose elements are all equal to zero, and we will denote with $M = D-Q_1-Q_2$ the number of common non-zero parts. \\

\begin{remark}
The setting in Eq.~\eqref{eq:comp_data_zeros} also includes the case where only the first $n_y$ observations have structural zeros, while the remaining observations are complete, i.e. $Q_2=0$.
\end{remark}

In the presence of structural zeros, logratio transformations cannot be applied to the data, and thus Aitchison's approach based on performing a PCA on the logratio-transformed data cannot be used. The currently available methods to tackle this issue can be grouped into three approaches. 

The first approach consists in the replacement of the zeros by an arbitrary small value. This procedure is very easy and effective, but, as argued in \citet[p.~57-58]{greenacre2018}, 
regardless of the zero-replacement strategy employed, the introduction of new values modifies the row totals of the dataset, thereby violating the constant-sum constraint. Then, it becomes necessary to apply the closure to the rows. However, these modifications in the data may significantly influence the results of the analyses.

A second approach suggests dropping out the zeros and performing the analyses with the remaining subcompositions with non-zero parts. This approach is also very easy to implement, but it has a relevant drawback: it is evident that the removal of all the null parts affects the amount of information conveyed by the data. In some cases, this loss of information can be significant, making the results of the analyses unreliable. 

Finally, the third approach consists in applying a different transformation, avoiding the logratios (see, for example, \citet{scealy2015}, \citet{lu2024} and the references therein). Among the transformations described in the literature, introduced, for example, in \cite{tsagris2011} or \cite{greenacre2024}, perhaps the most common one is the square root. The calculation of the square root of each part can be easily executed also in case of zero values, and it has the appealing peculiarity of modifying the dataset into a directional dataset, which can be handled by appropriate tools belonging to the field of directional statistics \citep{fisher1993, cuesta2009, lee2010, pewsey2013, pewsey2021, porro2024}.
Although this method can provide interesting and useful results, it does not adhere to Aitchison's original spirit \citep{alenazi2021}.



\section{Hypothesis testing for a common subspace}\label{sec:proptest}

Following the spirit of Aitchison's approach in presence of structural zeros, the procedure to reduce the dimension should consist in splitting the dataset $\mathcal{X}$ and performing two PCAs on the two sub-datasets $\mathcal Y$ and $\mathcal Z$ as defined in Eq. (\ref{eq:comp_data_zeros}). The rationale of this approach is that in the Aitchison framework, the two sub-datasets represent two different subpopulations each characterized by a different pattern of structural zeros. We follow this approach and try to overcome a relevant constraint: the observations can come from two subpopulations (identifiable by the structural zeros), but in many real cases, the investigation on what they share can be relevant. This means understanding whether the first (and most important) principal components of the two datasets span the same space. If this is true, it can be interpreted as evidence that both subpopulations actually share common characteristics.

Inspired by Aitchison's approach, we apply a PCA to the logratio-transformed data included in 
the sub-datasets $\mathcal{Y}$ and $\mathcal{Z}$. First, we compute the transformed datasets 

\begin{equation}\label{eq:ilr_datasets}
\transfDat{Y} = (\transf{y}_1, \dots,\transf{y}_{n_y})' \quad \mbox{and} \quad \transfDat{Z} = (\transf{z}_1, \dots,\transf{z}_{n_z})'
\end{equation}
where $\transf{y}_i = \ilr(\mathbf{y}_i)$ for all $i = 1, \dots, n_y$ and $\transf{z}_i = \ilr(\mathbf{z}_i)$ for all $i = 1, \dots, n_z$. 

\begin{remark} 
We observe that, by choosing the specific ilr transformation defined in Eq.~\eqref{eq:def_ilr}, the first $M-1$ components of the ilr-transformed data are not influenced by  the structural zeros and thus convey information only from the first $M$ common non-zero parts.  
\end{remark}

We then consider the sample covariance matrices $\widehat{\boldsymbol{\Omega}}_{\mathbf{Y}}$ and $\widehat{\boldsymbol{\Omega}}_{\mathbf{Z}}$ of $\transfDat{Y}$ and $\transfDat{Z}$, respectively, and we compute the eigenvalue decompositions
\begin{equation}\label{eq:PCA_transf_Y}
\widehat{\boldsymbol{\Omega}}_{\mathbf{Y}} = \widehat{\mathbf{U}} \,  \textrm{diag}(\widehat{\alpha}_1, \dots, \widehat{\alpha}_{D-Q_1-1}) \, \widehat{\mathbf{U}}' \quad \widehat{\alpha}_1 > \dots > \widehat{\alpha}_{D-Q_1-1}
\end{equation}
and 
\begin{equation}\label{eq:PCA_transf_Z}
\widehat{\boldsymbol{\Omega}}_{\mathbf{Z}} = \widehat{\mathbf{V}} \,  \textrm{diag}(\widehat{\beta}_1, \dots, \widehat{\beta}_{D-Q_2-1}) \, \widehat{\mathbf{V}}' \quad \widehat{\beta}_1 > \dots > \widehat{\beta}_{D-Q_2-1} \, .
\end{equation}

We observe that if the first $K$ principal components in Eqs.~\eqref{eq:PCA_transf_Y} and \eqref{eq:PCA_transf_Z} span the same subspace then this approach can be easily used to identify characteristics of the whole dataset $\mathcal{X}$. The main difficulty in comparing these components is that they belong to different vector spaces, because the columns of $\widehat{\mathbf{V}}$ and $\widehat{\mathbf{U}}$ are vectors of $\mathbb{R}^{D-Q_1-1}$ and $\mathbb{R}^{D-Q_2-1}$, respectively. This issue may be overcome by applying the canonical inclusions that transform $\widehat{\mathbf{u}}_i = (\widehat{u}_{1i}, \dots, \widehat{u}_{(D-Q_1-1)i})' \in \mathbb{R}^{D-Q_1-1}$ and $\widehat{\mathbf{v}}_i = (\widehat{v}_{1i}, \dots, \widehat{v}_{(D-Q_2-1)i})' \in \mathbb{R}^{D-Q_2-1}$ into, respectively
\begin{displaymath}
\begin{split}
& (\widehat{u}_{1i}, \dots, \widehat{u}_{(M-1)i}, \widehat{u}_{Mi}, \dots, \widehat{u}_{(D-Q_1-1)i}, \mathbf{0}_{1,Q_1})' \in \mathbb{R}^{D-1} \quad \mbox{and} \\
& (\widehat{v}_{1i}, \dots, \widehat{v}_{(M-1)i}, \mathbf{0}_{1,Q_2}, \widehat{v}_{Mi}, \dots, \widehat{v}_{(D-Q_2-1)i})' \in \mathbb{R}^{D-1} \, .
\end{split}
\end{displaymath}
A more compact notation of such immersions can be obtained by exploiting the following operators.
\begin{definition}\label{def:operators_iota}
Given the datasets $\mathcal{Y}$ and $\mathcal{Z}$ as in Eq.\eqref{eq:comp_data_zeros}, we define the linear operators $\iota_\mathcal{Y}:\mathbb{R}^{D-Q_1-1} \rightarrow \mathbb{R}^{D-1}$ and $\iota_\mathcal{Z}:\mathbb{R}^{D-Q_2-1} \rightarrow \mathbb{R}^{D-1}$ such that
\begin{equation}
\iota_\mathcal{Y} \mathbf{u} := \left(\begin{array}{cc}
\mathbf{I}_{M-1} &  \mathbf{0}_{M-1, Q_2} \\
\mathbf{0}_{Q_2, M-1} & \mathbf{I}_{Q_2} \\
\mathbf{0}_{Q_1, M-1} &
\mathbf{0}_{Q_1, Q_2} \end{array} \right) \mathbf{u}, \quad \quad \, \mathbf{u} \in \mathbb{R}^{D-Q_1-1} \, ,
\end{equation}
and 
\begin{equation}
\iota_\mathcal{Z} \mathbf{u} := \left(\begin{array}{cc}
\mathbf{I}_{M-1} &  \mathbf{0}_{M-1, Q_1} \\
\mathbf{0}_{Q_2, M-1} & \mathbf{0}_{Q_2,Q_1} \\
\mathbf{0}_{Q_1, M-1} &
\mathbf{I}_{Q_1} \end{array} \right) \mathbf{v}, \quad \quad \, \mathbf{v} \in \mathbb{R}^{D-Q_2-1} \, .
\end{equation}
\end{definition}

Motivated by this consideration, we introduce the following definition, which extends the concept of common principal component subspace proposed by \cite{schott1988} to settings where the two sub-datasets have distinct patterns of structural zeros.

For the sake of notational clarity, we first introduce the vectors, matrices, and notation that will be used in the subsequent definitions.

Let $\mathbf{Y}$ be a $(D-Q_1)$-dimensional random vector with covariance matrix $\boldsymbol{\Omega}^{(1)}$ and define $\mathbf{X}^{(1)} = \iota_\mathcal{Y}(\mathbf{Y})$. Similarly, let $\mathbf{Z}$ be a $(D-Q_2)$-dimensional random vector with covariance matrix $\boldsymbol{\Omega}^{(2)}$ and define $\mathbf{X}^{(2)} = \iota_\mathcal{Z}(\mathbf{Z})$, so that $\mathbf{X}^{(1)}$ and $\mathbf{X}^{(2)}$ have distinct patterns of structural zeros and share the first $M$ (non-zero) variables.

\begin{definition}\label{def:common_subspace}
In the above settings, fixed $K \in \{1, \dots, M-1\}$, we say that $\mathbf{X}^{(1)}$ and  $\mathbf{X}^{(2)}$ share a common principal component subspace of size $K$ if 
\begin{equation} \label{eq:com-subsp}
\spansb \left( \iota_{\mathcal{Y}} \mathbf{u}_1, \dots, \iota_{\mathcal{Y}} \mathbf{u}_K  \right) = \spansb \left(\iota_{\mathcal{Z}}\mathbf{v}_1, \dots, \iota_{\mathcal{Z}}\mathbf{v}_K\right)
\end{equation}
where $\mathbf{u}_1, \dots,  \mathbf{u}_K$ and $\mathbf{v}_1, \dots,  \mathbf{v}_K$ are the eigenvectors associated to the $K$ highest eigenvalues of $\boldsymbol{\Omega}^{(1)}$ and $\boldsymbol{\Omega}^{(2)}$, respectively.
\end{definition}

Moving to compositional data, we state the definition below.

\begin{definition}\label{def:common_subspace_comp}
Let $\mathbf{X}^{(1)}$ and  $\mathbf{X}^{(2)}$ be two compositional random vectors, and denote with $\boldsymbol{\Omega}^{(1)}$ and $\boldsymbol{\Omega}^{(2)}$ the covariance matrices of the pivot logratio transformations of the non-zero parts $\transfDat{Y}$ and $\transfDat{Z}$, respectively. Fixed $K \in \{ 1, \dots, M-1 \}$, we say that $\mathbf{X}^{(1)}$ and $\mathbf{X}^{(2)}$ share a common principal component subspace of size $K$ if the condition in Eq.~\eqref{eq:com-subsp} holds.
\end{definition}

In view of Definitions \ref{def:common_subspace} and \ref{def:common_subspace_comp}, the main objective of this paper is to develop a statistical procedure for testing the null hypothesis
\begin{equation}\label{eq:null_hp}
H_0: \spansb \left( \iota_{\mathcal{Y}} \mathbf{u}_1, \dots, \iota_{\mathcal{Y}} \mathbf{u}_K  \right) = \spansb \left(\iota_{\mathcal{Z}}\mathbf{v}_1, \dots, \iota_{\mathcal{Z}}\mathbf{v}_K\right)
\end{equation}
against the alternative hypothesis
\begin{equation}\label{eq:alt_hp}
H_1: \spansb \left( \iota_{\mathcal{Y}} \mathbf{u}_1, \dots, \iota_{\mathcal{Y}} \mathbf{u}_K  \right) \neq \spansb \left(\iota_{\mathcal{Z}}\mathbf{v}_1, \dots, \iota_{\mathcal{Z}}\mathbf{v}_K\right)
\end{equation}
where $\mathbf{u}_j \in \mathbb{R}^{D-Q_1-1}$ and $\mathbf{v}_j \in \mathbb{R}^{D-Q_2-1}$ denote the $j$-th principal component of $\transfDat{Y}$ and $\transfDat{Z}$, respectively.

Towards this end, inspired by \cite{schott1988}, we define the following test statistic, whose computation is summarized in Algorithm \ref{alg:test_sta_val}. 

\begin{definition}\label{def:test_statistics}
Let $\mathcal{X}$ be a compositional dataset with structural zeros as in Eq.~\eqref{eq:comp_data_zeros}. Assume that $\mathcal{Y}$ and $\mathcal{Z}$ are realizations of random samples $\mathbf{Y}_1, \dots, \mathbf{Y}_{n_y} \sim \mathbf{Y}$ and  $\mathbf{Z}_1, \dots, \mathbf{Z}_{n_z} \sim \mathbf{Z}$, as defined in Eq.~\eqref{eq:comp_data_zeros}. Fixed $K \in \{1, \dots, M-1 \}$ and denoted with $\widehat{\boldsymbol{\Omega}}_{\mathbf{Y}}$ and $\widehat{\boldsymbol{\Omega}}_{\mathbf{Z}}$ the sample covariance matrices of the data in pivot logratio coordinates, we define the test statistic
\begin{equation}\label{eq:def_T}
T = \sum_{i=1}^K \left( (n_y-1) \widehat{\alpha}_i + (n_z-1) \widehat{\beta}_i - \widehat{\gamma}_i  \right) \, ,
\end{equation}
where $\widehat{\alpha}_i$, $\widehat{\beta}_i$, and $\widehat{\gamma}_i$ denote the $i$-th largest eigenvalue of $\widehat{\boldsymbol{\Omega}}_{\mathbf{Y}}$, $\widehat{\boldsymbol{\Omega}}_{\mathbf{Z}}$, and 
\begin{equation}\label{eq:est_sum_matrix}
(n_y - 1) \iota_{\mathcal{Y}} \widehat{\boldsymbol{\Omega}}_{\mathbf{Y}} \iota_{\mathcal{Y}}' + (n_z - 1) \iota_{\mathcal{Z}} \widehat{\boldsymbol{\Omega}}_{\mathbf{Z}} \iota_{\mathcal{Z}}' \, , 
\end{equation}
respectively.
\end{definition}

\begin{remark}
Notice that with our approach the canonical immersion is applied to the (sample) covariance matrices and related eigenvalues and eigenvectors and not to the ilr-transformed data.  
\end{remark}

\begin{algorithm}[ht]\caption{Computation of the test statistic}\label{alg:test_sta_val}
\renewcommand{\algorithmicrequire}{\textbf{Input:}}
\renewcommand{\algorithmicensure}{\textbf{Output:}}
\begin{algorithmic}[1]
\REQUIRE{Compositional dataset with structural zeros as in Eq.~\eqref{eq:comp_data_zeros} \\
$K \in \{1, \dots, M-1\}$;}
\FOR{$i = 1, \dots n_y $}
\STATE{$\transf{y}_i \leftarrow \ilr(\mathbf{y}_i)$}
\ENDFOR
\FOR{$i = 1, \dots n_z $}
\STATE{$\transf{z}_i \leftarrow \ilr(\mathbf{z}_i)$}
\ENDFOR
\STATE{Assemble $\transfDat{Y} = \left(\transf{y}_1, \dots, \transf{y}_{n_y} \right)'$ and $\transfDat{Z} = \left(\transf{z}_1, \dots, \transf{z}_{n_z} \right)'$}
\STATE{Compute sample covariance matrices $\widehat{\boldsymbol{\Omega}}_{\mathbf{Y}}$ and $\widehat{\boldsymbol{\Omega}}_{\mathbf{Z}}$}
\STATE{Compute $K$ highest eigenvalues of $\widehat{\boldsymbol{\Omega}}_{\mathbf{Y}}$: $\widehat{\alpha}_1 \geq \dots \geq \widehat{\alpha}_K$}
\STATE{Compute $K$ highest eigenvalues of $\widehat{\boldsymbol{\Omega}}_{\mathbf{Z}}$: $\widehat{\beta}_1 \geq  \dots \geq \widehat{\beta}_K$}
\STATE{Compute $K$ highest eigenvalues of $(n_y - 1)  \iota_{\mathcal{Y}} \widehat{\boldsymbol{\Omega}}_{\mathbf{Y}} \iota_{\mathcal{Y}}'  + (n_z - 1) \iota_{\mathcal{Z}} \widehat{\boldsymbol{\Omega}}_{\mathbf{Z}} \iota_{\mathcal{Z}}'$: $\widehat{\gamma}_1 \geq  \dots \geq \widehat{\gamma}_K$}
\STATE{Compute $t = \sum_{j=1}^K \left((n_y -1) \widehat{\alpha}_j + (n_z -1) \widehat{\beta}_j - \widehat{\gamma}_j\right)$}
\ENSURE Test statistic $t$
\end{algorithmic}
\end{algorithm}


\subsection{Approximated test for normally distributed data}
\label{sec5}


Suppose that $\mathbf{y}_1, \dots,\mathbf{y}_{n_y}$ are realizations of a $(D-Q_1)$-part random composition $\mathbf{Y}$ having a normal distribution on the simplex $\mathbb{S}^{D-Q_1}$ with ilr-mean $\boldsymbol{\mu}_{\mathbf{Y}}$ and ilr-covariance matrix $\boldsymbol{\Omega}_{\mathbf{Y}}$, namely $\mathbf{Y} \sim \mathcal{N}_{\mathbb{S}^{D-Q_1}}(\boldsymbol{\mu}_{\mathbf{Y}},\boldsymbol{\Omega}_{\mathbf{Y}})$. It follows by definition \citep{egozcue2019} that the corresponding backward pivot logratio coordinates are realizations of a $(D-Q_1-1)$-dimensional random vector $\transf{Y} := \ilr(\mathbf{Y})$ with multivariate normal distribution $\mathcal{N}_{\mathbb{R}^{D-Q_1-1}}(\boldsymbol{\mu}_{\mathbf{Y}},\boldsymbol{\Omega}_{\mathbf{Y}})$. Analogously, suppose that $\mathbf{z}_1, \dots,\mathbf{z}_{n_z}$ are realizations of a $(D-Q_2)$-part random composition $\mathbf{Z} \sim \mathcal{N}_{\mathbb{S}^{D-Q_2}}(\boldsymbol{\mu}_{\mathbf{Z}},\boldsymbol{\Omega}_{\mathbf{Z}})$, and thus $\transf{Z} := \ilr(\mathbf{Z}) \sim \mathcal{N}_{\mathbb{R}^{D-Q_2-1}}(\boldsymbol{\mu}_{\mathbf{Z}},\boldsymbol{\Omega}_{\mathbf{Z}})$.

In this scenario, by exploiting that both $\transf{Y}$ and $\transf{Z}$ have a multivariate normal distribution, we are able to derive an analytical approximation for the distribution of the test statistic $T$ introduced in Definition \ref{def:test_statistics} under the null hypothesis that $\mathcal{Y}$ and $\mathcal{Z}$ share a common principal subspace, see Eq.~\eqref{eq:null_hp}. This result is summarized in the next theorem and is a generalization of the procedure proposed by \cite{schott1988} to the case in which the PCs of the two datasets 
belong to vector spaces with different dimensions.

\begin{theorem}\label{theo:distri_T}
Let us assume that the hypothesis of Definition \ref{def:test_statistics} hold and that $\mathbf{Y} \sim \mathcal{N}_{\mathbb{S}^{D-Q_1}}(\boldsymbol{\mu}_{\mathbf{Y}},\boldsymbol{\Omega}_{\mathbf{Y}})$ and $\mathbf{Z} \sim \mathcal{N}_{\mathbb{S}^{D-Q_2}}(\boldsymbol{\mu}_{\mathbf{Z}},\boldsymbol{\Omega}_{\mathbf{Z}})$. Then, the distribution of $T$ under the null hypothesis $H_0$ in Eq.~\eqref{eq:null_hp} can be approximated as 
\begin{equation}\label{eq:final_approx_T}
T \sim \frac{\sigma_T^2}{2 \mu_T} \chi^2\left(\left[\frac{2 \mu_T^2}{\sigma_T^2}\right]\right) \, ,
\end{equation}
where
\begin{equation}\label{eq:mu_t_theo}
\begin{split}
\mu_T & =  \sum_{i=1}^K \sum_{j=K+1}^{D-1} \Bigg\{  \frac{\widetilde{\alpha}_i \widetilde{\alpha}_j}{(\widetilde{\alpha}_i - \widetilde{\alpha}_j)} + \frac{\widetilde{\beta}_i \widetilde{\beta}_j}{(\widetilde{\beta}_i - \widetilde{\beta}_j)}  + \\
& - \frac{\sum_{h=1}^K \sum_{l = K+1}^{D-1} \left[ (n_y - 1) (u^*_{ih})^2(u^*_{jl})^2 \widetilde{\alpha}_{h}\widetilde{\alpha}_{l} + (n_z - 1) (v^*_{ih})^2(v^*_{jl})^2 \widetilde{\beta}_{h}\widetilde{\beta}_{l} \right]
}{(n_y + n_z - 2) (\psi_i - \psi_j)} \Bigg\}
\end{split}
\end{equation}
and
\begin{eqnarray}
\sigma^2_T & =& 2 \sum_{i=1}^K \sum_{j=K+1}^{D-1}\Bigg\{ \frac{\widetilde{\alpha}_i^2 \widetilde{\alpha}_j^2}{(\widetilde{\alpha}_i - \widetilde{\alpha}_j)^2} + \frac{ \widetilde{\beta}_i^2 \widetilde{\beta}_j^2}{(\widetilde{\beta}_i - \widetilde{\beta}_j)^2} - \frac{2}{(n_y + n_z - 2)(\psi_i 
- \psi_j)} \times \nonumber\\
& \times& \sum_{h=1}^K \sum_{l=K+1}^{D-1} \left[\frac{(n_y - 1) (\widetilde{\alpha}_{h}\widetilde{\alpha}_{l}u^*_{ih}u^*_{jl})^2 }{(\widetilde{\alpha}_h - \widetilde{\alpha}_l)} + \frac{(n_z - 1) (\widetilde{\beta}_{h}\widetilde{\beta}_{l} v^*_{ih}v^*_{jl})^2}{(\widetilde{\beta}_h - \widetilde{\beta}_l)}\right] +  \nonumber\\
& + &\sum_{h=1}^K \sum_{l=K+1}^{D-1} \frac{ W_{hl}^2}{(n_y+n_z-2)^2(\psi_i-\psi_j)(\psi_h - \psi_l)} \Bigg\} \, .\label{eq:sigma_T_theo}
\end{eqnarray}
with
\[
\begin{split}
W_{hl} & =(n_y-1)\left(\sum\limits_{s=1}^K \widetilde{\alpha}_s u_{is}^* u_{hs}^* \right)\left(\sum\limits_{t=K+1}^{D-1} \widetilde{\alpha}_t u_{jt}^* u_{l t}^* \right) \\
& + (n_z-1)\left(\sum\limits_{s=1}^K \widetilde{\beta}_s v_{is}^* v_{hs}^* \right)\left(\sum\limits_{t=K+1}^{D-1} \widetilde{\beta}_t v_{jt}^* v_{l t}^* \right).
\end{split}
\]

In Eqs.~\eqref{eq:mu_t_theo} and \eqref{eq:sigma_T_theo},  $(\widetilde{\alpha}_1, \dots, \widetilde{\alpha}_{D-1})' = \iota_\mathcal{Y}(\alpha_1, \dots, \alpha_{D-Q_1-1})'$ and $(\widetilde{\beta}_1, \dots, \widetilde{\beta}_{D-1})' = \iota_{\mathcal{Z}}(\beta_1, \dots, \beta_{D-Q_2-1})'$ are the zero-padded eigenvalues of  $\boldsymbol{\Omega}_{\mathbf{Y}}$ and $\boldsymbol{\Omega}_{\mathbf{Z}}$, respectively, while $\psi_1, \dots, \psi_{D-1}$ denote the eigenvalues of the pooled covariance matrix
\begin{equation}\label{eq:def_pooled_cov}
\boldsymbol{\Omega}_{pool} =(n_y+n_z-2)^{-1} \left[ (n_y - 1) \iota_{\mathcal{Y}}
\boldsymbol{\Omega}_{\mathbf{Y}} \iota_{\mathcal{Y}}' + (n_z - 1) \iota_{\mathcal{Z}}
\boldsymbol{\Omega}_{\mathbf{Z}} \iota_{\mathcal{Z}}' \right] \, .  
\end{equation}

Furthermore, denoted with $\mathbf{U} \in \mathbb{R}^{(D-Q_1-1) \times (D-Q_1-1)}$, $\mathbf{V} \in \mathbb{R}^{(D-Q_2-1) \times (D-Q_2-1)}$, and $\mathbf{K} \in \mathbb{R}^{(D-1) \times (D-1)}$ the matrix of the normalized eigenvectors of $\boldsymbol{\Omega}_{\mathbf{Y}}$, $\boldsymbol{\Omega}_{\mathbf{Z}}$, and $\boldsymbol{\Omega}_{pool}$, respectively, we defined $\mathbf{U}^* = \mathbf{K}' \left[ \iota_{\mathcal{Y}}
\mathbf{U} \iota_{\mathcal{Y}}' + (\mathbf{I}_{D-1} - \iota_{\mathcal{Y}} \iota_{\mathcal{Y}}')   \right]$ and $\mathbf{V}^* = \mathbf{K}' \left[ \iota_{\mathcal{Z}}
\mathbf{V} \iota_{\mathcal{Z}}' + (\mathbf{I}_{D-1} - \iota_{\mathcal{Z}} \iota_{\mathcal{Z}}')   \right]$.
\end{theorem}

The detailed proof of Theorem \ref{theo:distri_T} can be found in \ref{app_3}.

Eq.~\eqref{eq:mu_t_theo} and ~\eqref{eq:sigma_T_theo} describe the the parameters $\mu_T$ and $\sigma^2_T$ as functions of the eigenvectors and eigenvalues of the covariance matrices $\boldsymbol{\Omega}_{\mathbf{Y}}$ and $\boldsymbol{\Omega}_{\mathbf{Z}}$ that are usually unknown. To make Theorem \ref{theo:distri_T} applicable, we need, therefore, an estimate of such quantities. The unknown values can be estimated using the sample covariance matrices in pivot logratio coordinates as shown in the Algorithm \ref{alg:approx_mu_sigma}.

\begin{algorithm}[ht!]\caption{Approximation of the parameters of the null distribution for normally distributed data.}\label{alg:approx_mu_sigma}
\renewcommand{\algorithmicrequire}{\textbf{Input:}}
\renewcommand{\algorithmicensure}{\textbf{Output:}}
\begin{algorithmic}[1]
\REQUIRE  ilr-transformed datasets $\transfDat{Y}$ and $\transfDat{Z}$ as in Eq.~\eqref{eq:ilr_datasets}\\
\STATE{Compute $\widehat{\boldsymbol{\Omega}}_{\mathbf{Y}}$, eigenvectors $\widehat{\mathbf{U}}$ and eigenvalues $\widehat{\alpha}_1 \geq \dots \geq \widehat{\alpha}_{D-Q_1-1}$}\\
\STATE{Compute $\widehat{\boldsymbol{\Omega}}_{\mathbf{Z}}$, eigenvectors $\widehat{\mathbf{V}}$ and eigenvalues $\widehat{\beta}_1 \geq \dots \geq \widehat{\beta}_{D-Q_2-1}$}\\
\STATE{Compute matrix in Eq.~\eqref{eq:est_sum_matrix}, eigenvectors $\widehat{\mathbf{K}}$ and eigenvalues $\widehat{\gamma}_1 \geq \dots \geq \widehat{\gamma}_{D-1}$}\\
\STATE{Compute 
$$\widehat{\mathbf{U}}^* = 
\left[\begin{array}{cc} 
\mathbf{1}_K & \mathbf{0}_{K,(D-K-1)} \\
\mathbf{0}_{K,(D-K-1)} & \mathbf{1}_{D-K-1}\end{array} \right] \odot \left[\widehat{\mathbf{K}}'\left(\iota_{\mathcal{Y}}
\widehat{\mathbf{U}} \iota_{\mathcal{Y}}' + \mathbf{I}_{D-1} - \iota_{\mathcal{Y}} \iota_{\mathcal{Y}}' \right) \right]$$
and 
$$\widehat{\mathbf{V}}^* = 
\left[\begin{array}{cc} 
\mathbf{1}_K & \mathbf{0}_{K,(D-K-1)} \\
\mathbf{0}_{K,(D-K-1)} & \mathbf{1}_{D-K-1}\end{array} \right] \odot \left[\widehat{\mathbf{K}}'\left(\iota_{\mathcal{Z}}
\widehat{\mathbf{V}} \iota_{\mathcal{Z}}' + \mathbf{I}_{D-1} - \iota_{\mathcal{Z}} \iota_{\mathcal{Z}}' \right) \right]$$
where $\mathbf{1}_K$ and $\mathbf{1}_{D-K-1}$ are matrices of size $K \times K$ and $(D-K-1) \times (D-K-1)$, respectively, whose elements are equal to 1, and $\odot$ is the Hadamard pointwise product.}
\STATE{Approximate $\widehat{\mu}_T$ and $\widehat{\sigma}^2_T$ through Eq.~\eqref{eq:mu_t_theo} and ~\eqref{eq:sigma_T_theo} by replacing: \\
$\widetilde{\alpha}_i \rightarrow (\iota_{\mathcal{Y}}\widehat{\alpha})_i$; $\widetilde{\beta}_i \rightarrow (\iota_{\mathcal{Z}}\widehat{\beta})_i$; $\psi_i \rightarrow \widehat{\gamma}_i/(n_y+n_z-2)$; $u_{ij}^* \rightarrow \widehat{u}_{ij}^*$; $v_{ij}^* \rightarrow \widehat{v}_{ij}^*$}
\ENSURE $\widehat{\mu}_T$, $\widehat{\sigma}^2_T$
\end{algorithmic}
\end{algorithm}

\break

\break
\subsection{Nonparametric bootstrap test}
\label{sec:bootstrap}

The adapted Schott's procedure described in the previous section allows an analytic approximation of the null distribution of the test statistic $T$, provided that data are normally distributed on the simplex. Here we discuss a nonparametric approach to be applied when such an approximation does not hold. Since the presence of structural zeros prevents the permutation of observations between the two datasets $\mathcal{Y}$ and $\mathcal{Z}$, we base our approach on the bootstrap method described in Algorithm \ref{alg:boot}.

The Monte Carlo approximation of the distribution of the test statistic under the null hypothesis, and thus the approximation of the p-value, is performed as follows. We generate pairs of datasets $(\mathcal{Y}^b, \mathcal{Z}^b)$ that satisfy the null hypothesis in Eq.~\eqref{eq:null_hp}, i.e. share a common subspace of dimension $K$, even though the corresponding principal components are different. To this end, we first rotate the dataset $\transfDat{Y}$ so that the last $Q_2$ element of the first $K$ eigenvectors of the corresponding sample covariance matrix are zero. Specifically, we define the rotated dataset as $\transfDat{Y}^{boot} = \transfDat{Y} \boldsymbol{\mathcal{Q}}$, where $\boldsymbol{\mathcal{Q}}$ comes from the QR decomposition of the first $K$ columns of $\widehat{\mathbf{U}}$, that is
\begin{equation}\label{eq:QR_U1_K}
\left[
\begin{array}{ccc}
\widehat{\mathbf{u}}_1, \dots, \widehat{\mathbf{u}}_K
\end{array} \right] = \boldsymbol{\mathcal{Q}} \left[ \begin{array}{c} \boldsymbol{\mathcal{R}} \\ \mathbf{0}_{(D-Q_1-1-K),K}
\end{array} \right] \; .
\end{equation}
We then define $\transfDat{Z}^{boot}$ by applying to the dataset $\transfDat{Z}$ the rotation
\begin{equation}\label{eq:rot_Z_boot}
\mathbf{R}^{boot}_3 = \left[ \begin{array}{cc}
\boldsymbol{\mathcal{R}} & \mathbf{0}_{K,(D_2-K)} \\
\mathbf{0}_{(D_2-K),K} & \mathbf{I}_{(D_2-K)}
\end{array}  \right] \, \left[ \begin{array}{cc}
\mathbf{R}^{boot}_1 & \mathbf{0}_{K,(D_2-K)} \\
\mathbf{0}_{(D_2-K),K} & \mathbf{R}^{boot}_2
\end{array}  \right] \widehat{\mathbf{V}}' \; , 
\end{equation}
where $\mathbf{R}^{boot}_1$ and $\mathbf{R}^{boot}_2$ are random rotation matrices of size $K \times K$ and $(D-K-1) \times (D-K-1)$, respectively, and $D_2 = D - Q_2 - 1$. The rotation $\mathbf{R}^{boot}_3$ aligns the first $K$ principal components of $\transfDat{Z}^{boot}$ to those of $\transfDat{Y}^{boot}$ rotated according to $\mathbf{R}^{boot}_1$.

The bootstrap datasets $\mathcal{Y}^b$ and $\mathcal{Z}^b$ are formed by drawing with replacement $n_y$ samples from $\transfDat{Y}^{boot}$ and $n_z$ samples from $\transfDat{Z}^{boot}$, respectively.

\begin{algorithm}[ht]\caption{Nonparametric bootstrap test}\label{alg:boot}
\renewcommand{\algorithmicrequire}{\textbf{Input:}}
\renewcommand{\algorithmicensure}{\textbf{Output:}}
\begin{algorithmic}[1]
\REQUIRE ilr-transformed datasets $\transfDat{Y}$ and $\transfDat{Z}$ as in Eq.~\eqref{eq:ilr_datasets}\\
 $\widehat{\boldsymbol{\Omega}}_{\mathbf{Y}}$ and $\widehat{\boldsymbol{\Omega}}_{\mathbf{Z}}$, and corresponding eigenvectors $\widehat{\mathbf{U}}$ and $\widehat{\mathbf{V}}$ \\
 Sample value of the test statistic $t$ \\
 $n_{boot}$; $K$
 \STATE{Define $D_1 = D-Q_1-1$ and $D_2 = D-Q_2-1$}
\STATE{QR decomposition of the matrix formed by the first $K$ columns of $\widehat{\mathbf{U}}$ as in Eq.~\eqref{eq:QR_U1_K}}
\STATE{Randomly generate orthonormal matrices $\mathbf{R}^{boot}_1 \in \mathbb{R}^{K \times K}$ and $\mathbf{R}^{boot}_2 \in \mathbb{R}^{(D_2-K) \times (D_2-K)}$ and define $\mathbf{R}^{boot}_3$ as in Eq.~\eqref{eq:rot_Z_boot}}
\STATE{Define $\transfDat{Y}^{boot} = \transfDat{Y} \boldsymbol{\mathcal{Q}}$ and $\transfDat{Z}^{boot} = \transfDat{Z} (\mathbf{R}^{boot}_3)'$}
\FOR{$b = 1, \dots n_{boot} $}
\STATE{Define $\transfDat{Y}^b$ by drawing with repetition $n_y$ samples from 
$\transfDat{Y}^{boot}$}
\STATE{Define $\transfDat{Z}^b$ by drawing with repetition $n_z$ samples from 
$\transfDat{Z}^{boot}$}
\STATE{Compute the sample test statistic $t^b$ from $\transfDat{Y}^{boot}$ and $\transfDat{Z}^{boot}$ as in Algorithm \ref{alg:test_sta_val}}
\ENDFOR
\STATE{Approximate the p-value $p^{boot} = \frac{\# \left\{ t^b \geq t \right\}}{n_{boot}}$}
\ENSURE $p^{boot}$
\end{algorithmic}
\end{algorithm}


\section{Simulation study}
\label{sec:validationsim}

In this section we use simulated data to demonstrate the validity of the technique introduced in the previous sections. Towards this end, we considered datasets defined as in Eq.~\eqref{eq:comp_data_zeros} and we set $D=11$, $Q_1=2$, and $Q_2=3$. A preliminary numerical study for the parametric test in the simpler case of normally distributed data with structural zeros only in one sub-dataset (i.e. $Q_2=0$) can be found in \cite{porro2026}. We then set $K=2$ and we tested the null hypothesis in Eq.~\eqref{eq:null_hp} in three different scenarios.

\begin{itemize}
\item[\textit{S1}.] We simulated data under the null hypothesis by assuming that the first two PCs for the ilr-transformed datasets $\transfDat{Y}$ and $\transfDat{Z}$ span the same subspace. To this end we define the covariance matrices
\begin{equation}
\boldsymbol{\Omega}_{\mathbf{Y}} = \mathbf{U}  \textrm{diag}(\alpha_1, \dots, \alpha_{8}) \, \mathbf{U}'
\end{equation}
\begin{equation}
\boldsymbol{\Omega}_{\mathbf{Z}} = \mathbf{V}  \textrm{diag}(\beta_1, \dots, \beta_{7}) \, \mathbf{V}'
\end{equation}
where 
\begin{equation}
\mathbf{U} =  \left[ \begin{array}{cc}
\mathbf{U}_1 & \mathbf{0}_{5,3} \\
\mathbf{0}_{3,5} & \mathbf{U}_2 \;  
\end{array} \right]
\end{equation}
being $U_1 \in \mathbb{R}^{5 \times 5}$ and $U_2 \in \mathbb{R}^{3 \times 3}$ randomly generated orthonormal matrices \citep{stewart1980,Mezzadri2007}, and 
\begin{equation}
\mathbf{V} = \left[ \begin{array}{cc}
\mathbf{U}_1 & \mathbf{0}_{5,2} \\
\mathbf{0}_{2,5} & \mathbf{I}_{2}
\end{array} \right] 
\left[ \begin{array}{cc}
\mathbf{R}_1 & \mathbf{0}_{2,5} \\
\mathbf{0}_{5, 2} & \mathbf{R}_2
\end{array} \right]
\end{equation}
being $\mathbf{R}_1 \in \mathbb{R}^{2 \times 2}$ and $\mathbf{R}_2 \in \mathbb{R}^{5 \times 5}$ random rotation matrices.

\item[\textit{S2}.] Only the first PC is common for the ilr-transformed datasets $\transfDat{Y}$ and $\transfDat{Z}$. As in the previous scenario, we randomly sampled $\mathbf{U}_1$ and $\mathbf{U}_2$, while we defined 
\begin{equation}
\mathbf{V} = \left[ \begin{array}{cc}
\mathbf{U}_1 & \mathbf{0}_{5,2} \\
\mathbf{0}_{2,5} & \mathbf{I}_{2}
\end{array} \right] 
\left[ \begin{array}{cc}
1 & \mathbf{0}_{1,6} \\
\mathbf{0}_{6, 1} & \mathbf{R}
\end{array} \right]
\end{equation}
being $\mathbf{R} \in \mathbb{R}^{6\times6}$ a random rotation matrix.

\item[\textit{S3}.] The ilr-transformed datasets $\transfDat{Y}$ and $\transfDat{Z}$ do not share any common structure. In this case we defined $\mathbf{U}$ and $\mathbf{V}$ as independent randomly generated orthonormal matrices.
\end{itemize}

In each scenario, we sample the ilr-transformed data from three different families of probability distributions:

\begin{enumerate}
\item Zero-mean multivariate normal distributions with covariance matrices $\boldsymbol{\Omega}_{\mathbf{Y}}$ and $\boldsymbol{\Omega}_{\mathbf{Z}}$;

\item Zero-mean multivariate Student's $t$-distribution with scale matrices $\frac{\nu - 2}{\nu}\boldsymbol{\Omega}_{\mathbf{Y}}$ and $\frac{\nu - 2}{\nu}\boldsymbol{\Omega}_{\mathbf{Z}}$, where $\nu$ is the number of degrees of freedom. In the simulation below we set $\nu \in \{4, 8, 40\}$;

\item Uniform distributions on the hypercubes $[-\sqrt{3}, \sqrt{3}]^{8}$ and $[-\sqrt{3}, \sqrt{3}]^{7}$ rotated through the Cholesky decomposition of $\boldsymbol{\Omega}_{\mathbf{Y}}$ and $\boldsymbol{\Omega}_{\mathbf{Z}}$.
\end{enumerate}

With the choices above we obtain perfectly comparable results, since all the initial marginal distributions are standardized.\\

For each scenario and each distribution, we carried out 1000 simulations testing different sample sizes, namely $n_y \in \{20, 60, 100 \}$ and $n_z \in \{20, 60, 100 \}$. The eigenvalues of the covariance matrices were set equal to $(\alpha_1, \dots, \alpha_8) = (10, 9, 1, 1, 0.5, 0.4, 0.1, 0.01)$ and $(\beta_1, \dots,  \beta_7) = (6, 5, 1, 0.9, 0.3, 0.1, 0.02)$. Fixed $K=2$, we tested the null hypothesis in Eq.~\eqref{eq:null_hp} against the alternative hypothesis in Eq.~\eqref{eq:alt_hp} through the three procedures described in Section~\ref{sec:proptest}, i.e. by approximating the null distribution of the test statistic using Schott's formula with the true covariance matrices (\textit{Schott theo.}) and with the sample ones (\textit{Schott est.}) and using the bootstrap procedure described in Algorithm \ref{alg:boot} (\textit{Bootstrap}). In the \textit{Bootstrap} test, $n_{boot}=2000$ bootstrap samples were drawn. Of course, the \textit{Schott theo.} results can only be obtained in a simulation framework, since the true covariance matrix is not known in real data analyses. However, in this section we present also the \textit{Schott theo.} results, as some comparisons between \textit{Schott theo.} and \textit{Schott est.} are worth discussing.

To compare the three tests, we set their nominal level equal to $5\%$ and we estimated the probability of rejecting $H_0$ that corresponds to the probability of the type I error in Scenario \textit{S1} and with the power of the test in scenarios \textit{S2} and \textit{S3}.

\begin{table}[ht]
\centering
\begin{small}
\hspace*{-3cm}
\begin{tabular}{|c|c|ccccccccc|}
\hline
& & \multicolumn{9}{c|}{$(n_y, n_z)$} \\
  \hline
&  & (20,20) & (20,60) & (20,100) & (60,20) & (60,60) & (60,100) & (100,20) & (100,60) & (100,100) \\ 
  \hline
Gaussian & Schott theo. & 0.055 & 0.057 & 0.041 & 0.059 & 0.050 & 0.061 & 0.069 & 0.057 & 0.066 \\ 
 & Schott est. & 0.071 & 0.085 & 0.067 & 0.099 & 0.062 & 0.070 & 0.098 & 0.060 & 0.072 \\ 
 & Bootstrap & 0.068 & 0.088 & 0.068 & 0.104 & 0.055 & 0.062 & 0.095 & 0.061 & 0.070 \\ 
\hline
Student's $t$ & Schott theo. & 0.068 & 0.063 & 0.092 & 0.083 & 0.081 & 0.063 & 0.069 & 0.071 & 0.071 \\ 
 (40 df) & Schott est. & 0.084 & 0.080 & 0.117 & 0.114 & 0.083 & 0.068 & 0.095 & 0.083 & 0.071 \\ 
 & Bootstrap & 0.070 & 0.074 & 0.114 & 0.100 & 0.070 & 0.051 & 0.098 & 0.078 & 0.064 \\
 \hline
Student's $t$ & Schott theo. & 0.150 & 0.197 & 0.166 & 0.129 & 0.170 & 0.201 & 0.160 & 0.178 & 0.194 \\ 
(8 df) &  Schott est. & 0.192 & 0.209 & 0.216 & 0.191 & 0.189 & 0.190 & 0.208 & 0.213 & 0.192 \\ 
 & Bootstrap & 0.106 & 0.119 & 0.096 & 0.120 & 0.082 & 0.082 & 0.136 & 0.082 & 0.072 \\ 
\hline
Student's $t$ &Schott theo. & 0.288 & 0.365 & 0.347 & 0.276 & 0.456 & 0.481 & 0.323 & 0.443 & 0.531 \\ 
(4 df) &  Schott est. & 0.346 & 0.435 & 0.452 & 0.408 & 0.508 & 0.512 & 0.433 & 0.486 & 0.572 \\ 
  & Bootstrap & 0.169 & 0.187 & 0.191 & 0.217 & 0.154 & 0.127 & 0.243 & 0.158 & 0.156 \\ 
\hline
Uniform & Schott theo. & 0.036 & 0.025 & 0.037 & 0.035 & 0.040 & 0.023 & 0.027 & 0.036 & 0.030 \\ 
 & Schott est. & 0.049 & 0.037 & 0.065 & 0.065 & 0.050 & 0.025 & 0.044 & 0.035 & 0.034 \\ 
 & Bootstrap & 0.060 & 0.058 & 0.079 & 0.083 & 0.066 & 0.046 & 0.063 & 0.062 & 0.060 \\ 
\hline
\end{tabular}
\hspace*{-1cm}
\end{small}
\caption{Estimated probability of the type I error for varying sample sizes and sampling probability. Data were simulated according to the null hypothesis (Scenario \textit{S1}). The nominal value of tests is 5\%.}\label{tab:type_I_error}
\end{table}

\begin{table}[th]
\centering
\begin{small}
\hspace*{-3cm}
\begin{tabular}{|c|c|ccccccccc|}
  \hline
  & & \multicolumn{9}{c|}{$(n_y, n_z)$} \\
  \hline
 & & (20,20) & (20,60) & (20,100) & (60,20) & (60,60) & (60,100) & (100,20) & (100,60) & (100,100) \\ 
  \hline
  Gaussian & Schott theo. & 0.995 & 0.998 & 0.999 & 0.997 & 1.000 & 1.000 & 0.996 & 1.000 & 1.000 \\ 
 & Schott est. & 0.986 & 0.996 & 0.998 & 0.992 & 1.000 & 1.000 & 0.991 & 1.000 & 1.000 \\ 
 & Bootstrap & 0.987 & 0.996 & 0.998 & 0.995 & 1.000 & 1.000 & 0.991 & 1.000 & 1.000 \\
 \hline
 Student's $t$ & Schott theo. & 0.997 & 1.000 & 1.000 & 0.997 & 1.000 & 0.996 & 0.999 & 1.000 & 1.000 \\ 
 (40 df) & Schott est. & 0.987 & 0.999 & 0.999 & 0.989 & 1.000 & 1.000 & 0.995 & 1.000 & 1.000 \\ 
  & Bootstrap & 0.988 & 0.999 & 0.999 & 0.990 & 1.000 & 1.000 & 0.994 & 1.000 & 1.000 \\ 
  \hline 
  Student's $t$ & Schott theo. & 0.994 & 1.000 & 1.000 & 0.998 & 1.000 & 0.998 & 0.996 & 1.000 & 1.000 \\ 
  (8 df) & Schott est. & 0.983 & 0.995 & 0.999 & 0.991 & 1.000 & 1.000 & 0.989 & 1.000 & 1.000 \\ 
  & Bootstrap & 0.981 & 0.995 & 0.998 & 0.991 & 0.999 & 1.000 & 0.986 & 1.000 & 1.000 \\ 
  \hline
  Student's $t$ & Schott theo. & 0.993 & 0.996 & 0.999 & 0.997 & 0.999 & 1.000 & 0.993 & 0.999 & 1.000 \\ 
  (4 df) & Schott est. & 0.968 & 0.996 & 0.996 & 0.988 & 0.997 & 1.000 & 0.981 & 1.000 & 1.000 \\ 
  & Bootstrap & 0.968 & 0.989 & 0.989 & 0.983 & 0.993 & 0.996 & 0.979 & 0.998 & 0.998 \\ 
  \hline
  Uniform & Schott theo. & 0.996 & 1.000 & 1.000 & 0.999 & 1.000 & 0.998 & 0.999 & 1.000 & 1.000 \\ 
  & Schott est. & 0.988 & 0.996 & 1.000 & 0.994 & 1.000 & 0.999 & 0.995 & 1.000 & 1.000 \\ 
  & Bootstrap & 0.993 & 1.000 & 1.000 & 0.996 & 1.000 & 1.000 & 0.998 & 1.000 & 1.000 \\
\hline
\end{tabular}
\end{small}
\hspace*{-3cm}
\caption{Estimated power of the tests for varying sample sizes and sampling probability. Data were simulated according to the alternative hypothesis as described in scenario \textit{S2}. The nominal value of tests is 5\%.}\label{tab:power_K1}
\end{table}

\begin{table}[th]
\centering
\hspace*{-3cm}
\begin{small}
\begin{tabular}{|c|c|ccccccccc|}
  \hline
  & & \multicolumn{9}{c|}{$(n_y, n_z)$} \\
  \hline
 & & (20,20) & (20,60) & (20,100) & (60,20) & (60,60) & (60,100) & (100,20) & (100,60) & (100,100) \\ 
  \hline
Gaussian &Schott theo. & 0.999 & 1.000 & 1.000 & 0.999 & 1.000 & 0.999 & 0.999 & 1.000 & 1.000 \\ 
 & Schott est. & 0.991 & 0.995 & 1.000 & 0.996 & 1.000 & 0.998 & 0.995 & 1.000 & 1.000 \\ 
 & Bootstrap & 0.997 & 1.000 & 1.000 & 0.998 & 1.000 & 1.000 & 0.999 & 1.000 & 1.000 \\ 
 \hline
  Student's $t$ & Schott theo. & 0.995 & 0.999 & 1.000 & 1.000 & 1.000 & 0.997 & 1.000 & 1.000 & 1.000 \\ 
 (40 df) & Schott est. & 0.990 & 0.997 & 1.000 & 0.996 & 1.000 & 0.996 & 0.998 & 1.000 & 1.000 \\ 
  & Bootstrap & 0.995 & 0.998 & 1.000 & 0.997 & 1.000 & 1.000 & 0.999 & 1.000 & 1.000 \\ 
  \hline
  Student's $t$ & Schott theo. & 0.996 & 1.000 & 1.000 & 0.999 & 1.000 & 0.993 & 0.999 & 1.000 & 1.000 \\ 
  (8 df) & Schott est. & 0.993 & 0.997 & 1.000 & 0.997 & 0.999 & 0.996 & 0.999 & 1.000 & 1.000 \\ 
 &  Bootstrap & 0.994 & 1.000 & 1.000 & 0.996 & 1.000 & 1.000 & 0.997 & 1.000 & 1.000 \\
 \hline
 Student's $t$ & Schott theo. & 0.995 & 1.000 & 0.999 & 0.998 & 1.000 & 0.997 & 0.999 & 1.000 & 1.000 \\ 
 (4 df) &  Schott est. & 0.979 & 0.996 & 0.997 & 0.994 & 0.997 & 0.999 & 0.995 & 1.000 & 0.999 \\ 
 & Bootstrap & 0.978 & 0.998 & 0.996 & 0.993 & 0.998 & 0.998 & 0.991 & 1.000 & 1.000 \\
 \hline
 Uniform & Schott theo. & 1.000 & 1.000 & 1.000 & 0.999 & 1.000 & 0.997 & 1.000 & 1.000 & 1.000 \\ 
 & Schott est. & 0.994 & 1.000 & 1.000 & 0.998 & 0.999 & 0.998 & 1.000 & 1.000 & 1.000 \\ 
 & Bootstrap & 1.000 & 1.000 & 1.000 & 0.999 & 1.000 & 1.000 & 1.000 & 1.000 & 1.000 \\
   \hline
\end{tabular}
\end{small}
\hspace*{-3cm}
\caption{Estimated power of the tests for varying sample sizes and sampling probability. Data were simulated according to the alternative hypothesis as described in scenario \textit{S3}. The nominal value of tests is 5\%.}\label{tab:power_K0}
\end{table}

Under the Gaussian assumption for the simulated data, \textit{Schott theo.} exhibits in most cases the type I error rate closest to the nominal value (Table \ref{tab:type_I_error}) and the highest power (Tables \ref{tab:power_K1} and \ref{tab:power_K0}). Instead, \textit{Schott est.} and \textit{Bootstrap} provide comparable results, although \textit{Bootstrap} outperforms \textit{Schott est.} for small sample sizes.

When data are sampled from multivariate Student's $t$-distributions all tests tend to be too liberal. However, in this scenario the \textit{Bootstrap} test shows in general a lower probability of type I error, see Table \ref{tab:type_I_error}. To highlight the most extreme example, when the number of degrees of freedom is 4, the estimated probability of type I error ranges from 0.127 to 0.243 for the \textit{Bootstrap} test, from 0.276 to 0.531 for \textit{Schott theo.}, and from 0.346 to 0.572 for \textit{Schott est.}. These values show that in case of fat tails the asymptotic procedure based on the Schott's scheme is completely unreliable in terms of bias. Moreover, the power of the \textit{Bootstrap} test is in general higher than the power of \textit{Schott est.}, see Tables \ref{tab:power_K1} and \ref{tab:power_K0}.

When a multivariate Uniform distribution is used to simulate data, the tests based on Schott's approximation appear too conservative, as their type I error rates fall below the nominal level 0.05, see Table \ref{tab:type_I_error}. Instead the type I error rate of the \textit{Bootstrap} test ranges between 0.046 and 0.083. In this scenario, all the considered tests show a power close to 1, except \textit{Schott est.}, which shows lower power at the smallest sample sizes.



\section{Real-data example}\label{sec:validationreal}

To illustrate the theory on a real-world dataset we consider here data coming from the Human Microbiome Project (HMP) available through the \texttt{Bioconductor} package \texttt{HMP16SData} \citep{HMP16Spackage,schiffer}, which was developed to provide efficient access to HMP 16S rRNA (16S ribosomal RNA) sequencing data within the \texttt{R/Bioconductor} environment. These data are widely used as benchmark data in microbiome studies and, more generally, in computational statistics applications, especially in the framework of species-abundance analysis. The analyzed data consist of 16S rRNA sequencing profiles, represented as an OTU (Operational Taxonomic Unit)-by-sample count matrix and accompanied by taxonomic annotation. Starting from the raw counts, OTUs were aggregated at the phylum level using the available taxonomic information. Samples with a sequencing depth below 1,000 total reads were removed in order to reduce the impact of poorly represented samples. Phyla not observed in the considered dataset were discarded and, when specified, unclassified sequences were excluded. The aggregated counts were then transformed into sample-specific relative compositions, yielding, for each sample, the proportional distribution of the detected bacterial phyla. In a subsequent filtering step, only phyla belonging to the native support of the dataset were retained, defined according to a minimum prevalence threshold of $5\%$ of samples. The resulting compositions were finally renormalized, yielding sample-by-phylum matrices directly comparable in the subsequent analyses.

Samples are further grouped according to the anatomical body site from which they were collected. A total of 18 sites are available: the \textit{Anterior Nares}, the \textit{Stool}, nine oral sites, four skin sites, and three sites from the urogenital tract. Based on this structure, the aim of the analysis is to assess, through pairwise comparisons between body sites, whether the microbial compositions aggregated at the phylum level share a common linear subspace. 


In particular, we illustrate here two comparisons: \textit{Right Antecubital Fossa} vs \textit{Anterior Nares} and \textit{Stool} vs \textit{Saliva}. As shown in Table \ref{tab:native_supports}, in the first case (Comparison 1) $M=10$ shared phyla belong to the native support of both sites, $Q_2=4$ phyla are structural zeros for the Anterior Nares, while $Q_1=0$. In the second case (Comparison 2) $M=7$ shared phyla are preset, while $Q_1=5$ and $Q_2=2$.

\begin{table}[ht]
\centering
\begin{small}
\begin{tabular}{|c|c|l|}
\hline 
\parbox[t]{2mm}{\multirow{7}{*}{\rotatebox[origin=c]{90}{Comparison 1}}} & \multirow{4}{*}{\shortstack{R.A. Fossa \\ ($n_y=101)$}} & \textbf{Firmicutes, Proteobacteria, Bacteroidetes, Cyanobacteria,} \\
& & \textbf{Actinobacteria, Fusobacteria, Tenericutes, Spirochaetes,} \\ 
& & \textbf{Thermi, Acidobacteria}, Chloroflexi, Verrucomicrobia, \\
& & Euryarchaeota, Gemmatimonadetes\\
\cline{2-3}
& \multirow{3}{*}{\shortstack{Ant. Nares \\ ($n_z = 240$)}} & \textbf{Firmicutes, Proteobacteria, Bacteroidetes, Cyanobacteria,} \\
& & \textbf{Actinobacteria, Fusobacteria, Tenericutes, Spirochaetes,} \\
& & \textbf{Thermi, Acidobacteria} \\
\hline 
\hline 
\parbox[t]{2mm}{\multirow{6}{*}{\rotatebox[origin=c]{90}{Comparison 2}}}
& \multirow{3}{*}{\shortstack{Stool \\ ($n_y = 305$)}} & \textbf{Firmicutes, Proteobacteria, Bacteroidetes, Cyanobacteria,} \\ 
& & \textbf{Actinobacteria, Fusobacteria, Tenericutes,} Verrucomicrobia, \\ & & Lentisphaerae \\
\cline{2-3}
& \multirow{3}{*}{\shortstack{Saliva \\ $(n_z = 273)$ }} &  \textbf{Firmicutes, Proteobacteria, Bacteroidetes, Cyanobacteria,} \\
& & \textbf{Actinobacteria, Fusobacteria, Tenericutes,} Spirochaetes, \\
& & GN02, Synergistetes, TM7, SR1 \\
\hline
\end{tabular}
\end{small}
\caption{Phyla within the native support of the considered body sites. For each performed comparison, phyla shared by both sites are highlighted in bold. Sample sizes are reported in parenthesis under the sites names.\\
R.A. Fossa = Right Antecubital Fossa. Ant. Nares = Anterior Nares}\label{tab:native_supports}
\end{table}

\begin{figure}
\centering
\includegraphics[width = \textwidth]{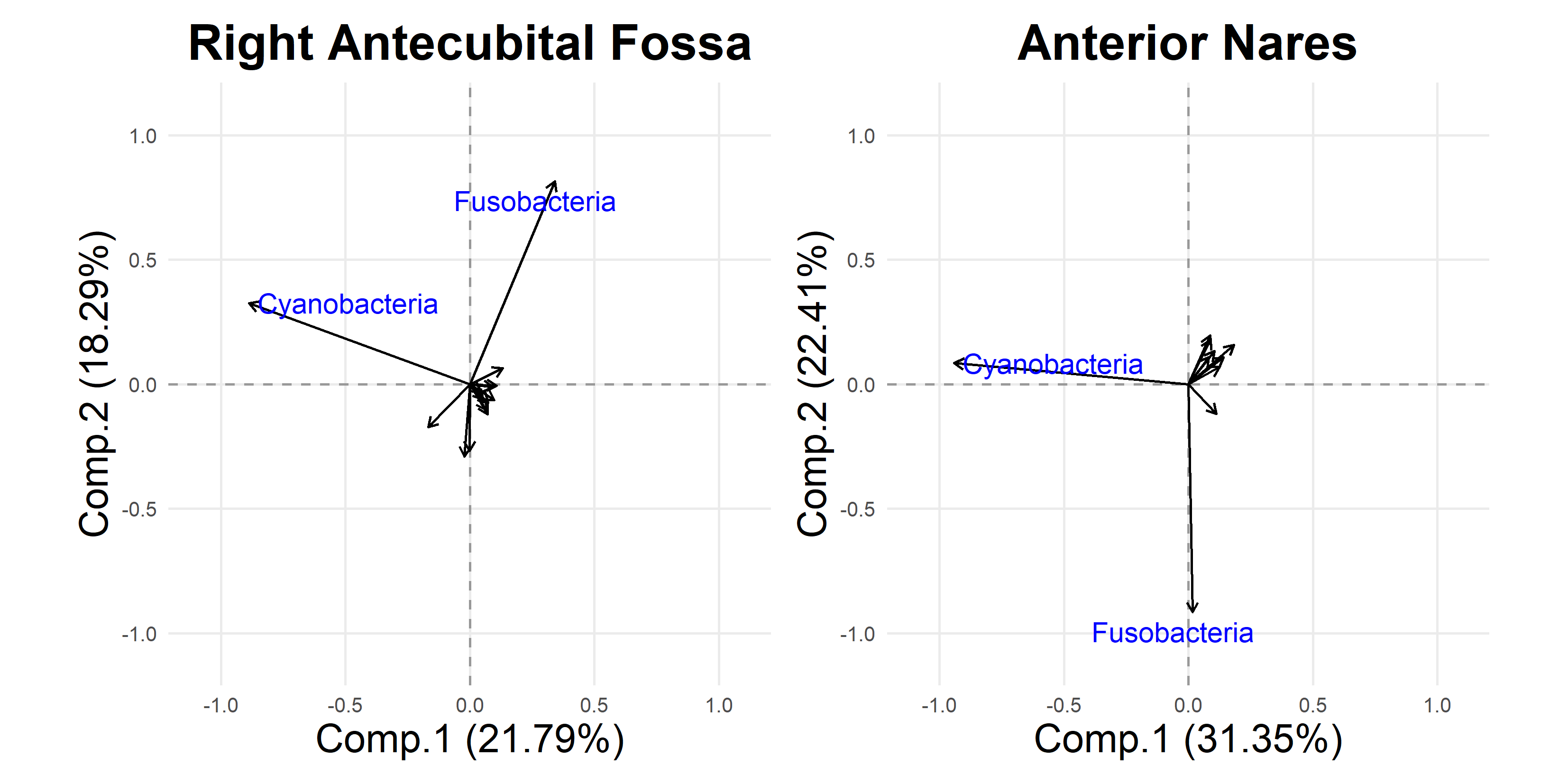}
\caption{Relative variation loading biplots showing the contribution of bacterial phyla to the first two Principal Components in the Right Antecubital Fossa (left) and in Anterior Nares (right)}\label{fig:ex_common_subspace}
\end{figure}

\begin{figure}
\centering
\includegraphics[width = \textwidth]{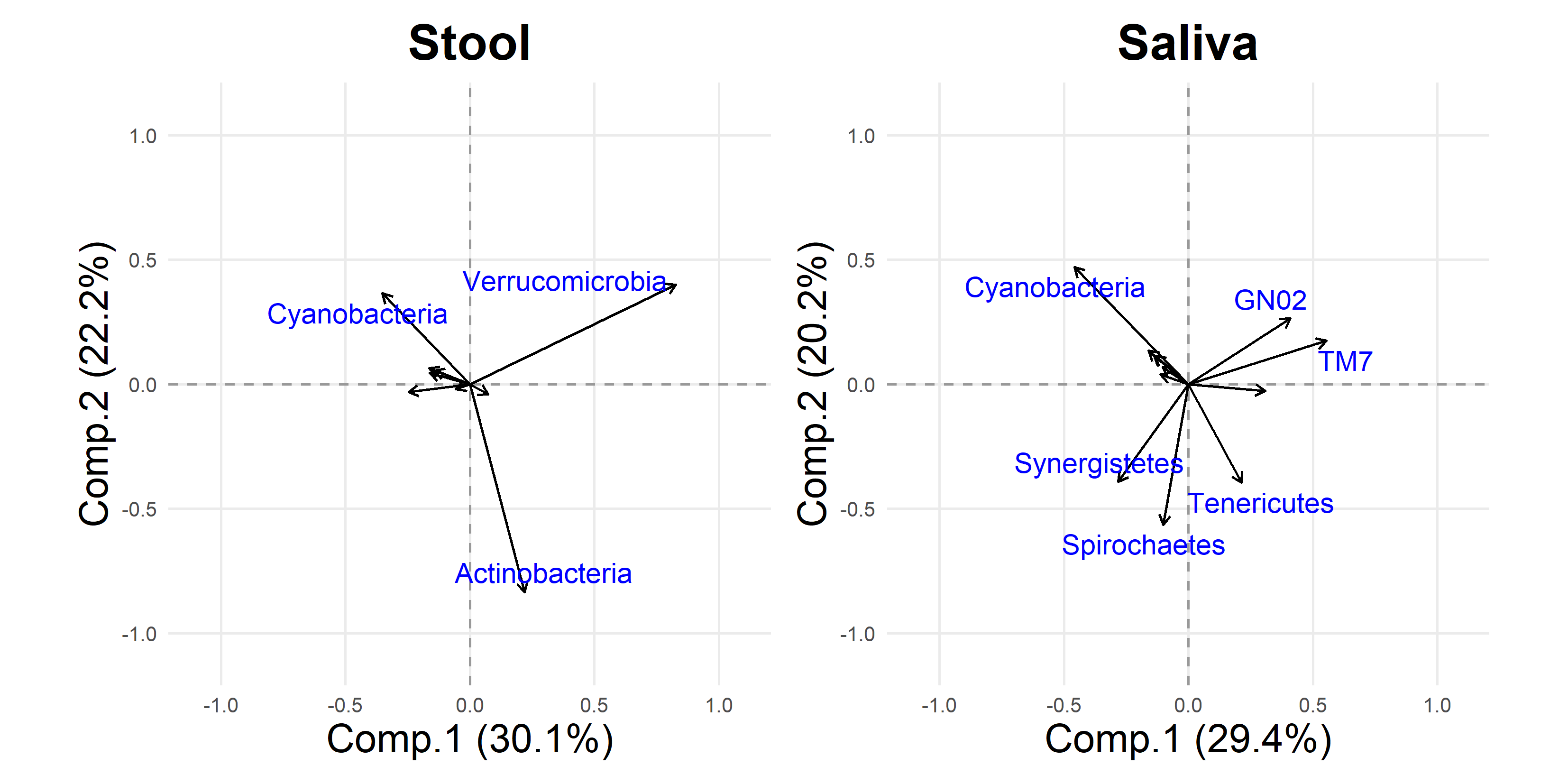}
\caption{Relative variation loading biplots showing the contribution of bacterial phyla to the first two Principal Components in Stool (left) and Saliva (right) samples.}\label{fig:ex_reject_common_subspace}
\end{figure}

For both comparisons, We tested the null hypothesis in Eq.~\eqref{eq:null_hp} with $K \in \left\{1, \dots, M \right\}$ and significance level equal to $0.05$. We used both, the modified Schott's test for normally distributed data described in Section \ref{sec5} and the bootstrap procedure described in Section \ref{sec:bootstrap}, with $n_{boot} = 2000$  bootstrap samples. When Right Antecubital Fossa and Anterior Nares are compared, the null hypothesis in Eq.~\eqref{eq:null_hp} is not rejected for $K=1$ (p-value modified Schott: $0.71$; bootstrap $0.86$) and $K=2$ (p-value modified Schott: $0.65$; bootstrap: $0.72$). As can be seen from the relative variation biplot \citep{aitchison2002} of the variable loadings in Fig. \ref{fig:ex_common_subspace} two are the main phyla contributing to this common subspace, namely the Cyanobacteria and the Fusobacteria. It is worth noting that both phyla show low relative abundance in both sites. However, their contribution to the first two principal components appears to be driven by their prevalence pattern: because these phyla are sparse, the logratio transformation produces high variance between the samples that contain them and those that do not. Whether this presence/absence pattern reflects a meaningful biological distinction between samples requires further investigation and is beyond the scope of this paper. Conversely, when Stool and Saliva are compared, the null hypothesis is rejected for all the considered $K$ (p-value lower then 0.001 for both modified Schott and bootstrap procedures) and the absence of any shared subspace is confirmed by the relative variation biplot in Fig. \ref{fig:ex_reject_common_subspace}.

\section{Discussion}\label{sec:discussion}

In this study, we introduced a statistical procedure for testing the existence of a common principal component subspace between two compositional datasets, both of which may contain structural zeros. Specifically, we adapted Schott's test \citep{schott1988} to handle compositional data that are normally distributed on the simplex, and we proposed a bootstrap-based test for the nonparametric case.

Our simulation results show that the bootstrap test performs comparably to the adapted Schott's test when the data are sampled from Gaussian distributions on the simplex. Conversely, in the vast majority of scenarios, the bootstrap approach outperforms Schott's test by achieving higher statistical power while maintaining the type I error rate closer to the nominal significance level. The real-data application, which compares the compositional structure of the human microbiome across different body sites, demonstrates that the proposed testing procedure provides meaningful and readily interpretable insights.

From a methodological point of view, both the proposed parametric and non-parametric tests require the size $K$ of the possibly common subspace to be fixed before testing for its existence and focus on the comparison between two datasets. Future effort will be devoted in designing a data-driven approach for choosing $K$ and in testing for a subspace common to three or more datasets. Furthermore, in this paper, we have shown how to exploit the relative variation biplots of the two datasets to visually inspect the variables contributing to the common subspace when the null hypothesis is not rejected. Future work will be devoted to developing a quantitative approach for actually determining the common subspace and to deepening the biological interpretation of the real data analysis.

\section*{Acknowledgments}

The research of F.P. and S.S. was partially support by the project ``DHEAL-COM - Digital Health Solutions in Community Medicine'' (CUP: D33C22001980001) under the Innovative Health Ecosystem (PNC)—National Recovery and Resilience Plan (NRRP) program funded by the Italian Ministry of Health.

F.P. is member of the INdAM-GNAMPA group, S.S. is member of the INdAM-GNCS group.


\appendix
\section{Null distribution of $T$ when data are normally distributed on the simplex.}
\label{app_proof_theo1}

\subsection{Auxiliary results}

For the proof of Theorem \ref{theo:distri_T} we make use of the following properties of the sample covariance matrices \citep[cfr.][]{bishop2018}. 
\begin{theorem}\label{theo:wish_1}
Let $\mathbf{X}_1, \dots, \mathbf{X}_n$ be $n$ i.i.d. $P$-dimensional random vectors following a multivariate normal distribution $\mathcal{N}_P(\boldsymbol{\mu}, \boldsymbol{\Omega})$. Denoted with $\overline{\mathbf{X}} = \frac{1}{n} \sum_{i=1}^n \mathbf{X}_i$ and with $\widehat{\boldsymbol{\Omega}} = \frac{1}{n-1} \sum_{i=1}^n \left(\mathbf{X}_i - \overline{\mathbf{X}}\right)\left(\mathbf{X}_i - \overline{\mathbf{X}}\right)'$ the corresponding sample mean and sample covariance matrix it holds
$$(n-1)\widehat{\boldsymbol{\Omega}} \sim \mathcal{W}_P(\boldsymbol{\Omega}, n-1) \, .$$
\end{theorem}

\begin{theorem}\label{theo:wish_2}
If $\mathbf{S} \sim \mathcal{W}_P(\boldsymbol{\Omega}, m)$ and $\mathbf{C} \in \mathbb{R}^{Q \times P}$ is a matrix with rank $P$, then $\mathbf{C}\mathbf{S}\mathbf{C}' \sim \mathcal{W}_P(\mathbf{C}\boldsymbol{\Omega}\mathbf{C}', m)$.
\end{theorem}

\begin{theorem}\label{theo:approx_eigen_covmat}
Assume $\mathbf{S} \sim \mathcal{W}_P(\diag(\delta_1, \dots, \delta_P), n)$ with $\delta_1 \geq \delta_2 \geq \dots \geq \delta_K > \delta_{K+1} \geq \dots \geq \delta_P$ and define $\mathbf{W}= \mathbf{S} - n\ \diag(\delta_1, \dots, \delta_P)$. Then, up to second order terms in the elements of $n^{-1}\mathbf{W}$, it holds
$$\sum_{i=1}^K \lambda_i(\mathbf{S}) \simeq n \left(\sum_{i=1}^K \delta_i + \sum_{i=1}^K \frac{w_{ii}}{n} + \sum_{i=1}^K \sum_{j=K+1}^{D-1}\frac{w_{ij}^2}{n^2 (\delta_i - \delta_j)}\right) \, $$
where $\lambda_i(\mathbf{S})$ denotes the $i-$th greatest eigenvalue of $\mathbf{S}$.
\end{theorem}

\subsection{Proof of Theorem \ref{theo:distri_T}}
\label{app_3}

\begin{proof} For the ease of exposition, henceforth we denote 
$$
\mathbf{U}^{\hookrightarrow} = \iota_{\mathcal{Y}}
\mathbf{U} \iota_{\mathcal{Y}}' + (\mathbf{I}_{D-1} - \iota_{\mathcal{Y}} \iota_{\mathcal{Y}}') \quad \quad 
\mathbf{V}^{\hookrightarrow} =  \iota_{\mathcal{Z}}
\mathbf{V} \iota_{\mathcal{Z}}' + (\mathbf{I}_{D-1} - \iota_{\mathcal{Z}} \iota_{\mathcal{Z}}') \; ,
$$
and given a matrix $\mathbf{M}$ we denote with $\mathbf{m}_i$ its $i$-th column and with $\trace(\mathbf{M})$ its trace.

We observe that if the null hypothesis $H_0$ in Eq.~\eqref{eq:null_hp} is true then there exist orthogonal matrices $\mathbf{R}_1 \in \mathbb{R}^{K \times K}$ and $\mathbf{R}_2 \in \mathbb{R}^{(D-K-1) \times (D-K-1)}$ such that $(\mathbf{u}^{\hookrightarrow}_1, \dots, \mathbf{u}^{\hookrightarrow}_K) = (\mathbf{v}^{\hookrightarrow}_1, \dots, \mathbf{v}^{\hookrightarrow}_K) \mathbf{R}_1$ and $(\mathbf{u}^{\hookrightarrow}_{K+1}, \dots, \mathbf{u}^{\hookrightarrow}_{D-1}) = (\mathbf{v}^{\hookrightarrow}_{K+1}, \dots, \mathbf{v}^{\hookrightarrow}_{D-1}) \mathbf{R}_2$. By exploiting these equalities into Eq.~\eqref{eq:def_pooled_cov} we can rewrite the pooled covariance matrix as
$$
\boldsymbol{\Omega}_{pool} = \mathbf{V}^{\hookrightarrow} \left[\begin{array}{cc} \boldsymbol{\Sigma}_1 & \mathbf{0}_{K,D-K-1} \\
\mathbf{0}_{D-K-1,K} & \boldsymbol{\Sigma}_2
\end{array} \right]  (\mathbf{V}^{\hookrightarrow})'
$$
where $\boldsymbol{\Sigma}_1  \in \mathbb{R}^{K \times K}$ and $\boldsymbol{\Sigma}_2  \in \mathbb{R}^{(D-K-1) \times (D-K-1)}$ are symmetric matrices defined as
$$
\boldsymbol{\Sigma}_1 = \mathbf{R}_1 \, \frac{n_y-1}{n_y+n_z -2} \diag(\widetilde{\alpha}_1, \dots, \widetilde{\alpha}_K) \mathbf{R}_1' + \frac{n_z -1}{n_y+n_z-2} \diag(\widetilde{\beta}_1, \dots, \widetilde{\beta}_K)
$$
$$
\boldsymbol{\Sigma}_2 = \mathbf{R}_2 \, \frac{n_y-1}{n_y+n_z -2} \diag(\widetilde{\alpha}_{K+1}, \dots, \widetilde{\alpha}_{D-1}) \mathbf{R}_2' + \frac{n_z -1}{n_y+n_z-2} \diag(\widetilde{\beta}_{K+1}, \dots, \widetilde{\beta}_{D-1})
$$
It follows that, under $H_0$, the following two properties hold:
\begin{itemize}
\item[(i)] the $K$ greatest eigenvalues of $\boldsymbol{\Omega}_{pool}$ correspond to the eigenvalues of $\boldsymbol{\Sigma}_1$, and hence
\begin{equation}\label{eq:rel_latentval_H0}
\sum_{i=1}^K \psi_i = \trace(\boldsymbol{\Sigma}_1) = \frac{n_y-1}{n_y+n_z-2} \sum_{i=1}^K \widetilde{\alpha}_i + \frac{n_z-1}{n_y+n_z-2} \sum_{i=1}^K \widetilde{\beta}_i; 
\end{equation}
\item[(ii)] the normalized eigenvectors of $\boldsymbol{\Omega}_{pool}$ are of the form $$\mathbf{K} = \mathbf{V}^{\hookrightarrow} \left[\begin{array}{cc} \mathbf{Q}_1 & \mathbf{0}_{K,D-K-1} \\
\mathbf{0}_{D-K-1,K} & \mathbf{Q}_2 \end{array} \right],$$
being $\mathbf{Q}_1$ and $\mathbf{Q}_2$ the matrices of the normalized eigenvectors of $\boldsymbol{\Sigma}_1$ and $\boldsymbol{\Sigma}_2$, respectively. As a consequence, both $\mathbf{V}^*$ and $\mathbf{U}^*$ are block diagonal matrices of the form
$$
\mathbf{V}^* = \left[\begin{array}{cc} \mathbf{V}^*_1 & \mathbf{0}_{K,D-K-1} \\
\mathbf{0}_{D-K-1,K} & \mathbf{V}^*_2 \end{array} \right] \quad \text{and} \quad \mathbf{U}^* = \left[\begin{array}{cc} \mathbf{U}^*_1 & \mathbf{0}_{K,D-K-1} \\
\mathbf{0}_{D-K-1,K} & \mathbf{U}^*_2 \end{array}  \right] \, .
$$
\end{itemize}
Since by hypothesis $\mathbf{y}_1, \dots, \mathbf{y}_{n_y}$ and $\mathbf{z}_1, \dots, \mathbf{z}_{n_z}$ are realizations of random compositions normally distributed on the simplex, the corresponding pivot logratio coordinates are realization of Gaussian random vectors and hence Theorem \ref{theo:wish_1} and Theorem \ref{theo:wish_2} can be applied to the corresponding sample covariance matrices. This leads to the following results:
\begin{equation}\label{eq:wishart_Y}
(n_y-1)\mathbf{U}'\widehat{\boldsymbol{\Omega}}_{\mathbf{Y}}\mathbf{U} \sim \mathcal{W}_{D-Q_1-1}\left(\diag(\alpha_1, \dots, \alpha_{D-Q_1-1}), n_y-1\right)
\end{equation}
and
\begin{equation}\label{eq:wishart_Z}
(n_z-1)\mathbf{V}'\widehat{\boldsymbol{\Omega}}_{\mathbf{Z}}\mathbf{V} \sim \mathcal{W}_{D-Q_2-1}\left(\diag(\beta_1, \dots, \beta_{D-Q_2-1}), n_z-1\right) \, .
\end{equation}
In order to apply Theorem \ref{theo:approx_eigen_covmat}, we define the matrix of size $(D-1) \times (D-1)$
\begin{equation}\label{eq:rel_errmat_A}
\begin{split}
\widetilde{\mathbf{A}}& := (n_y-1) \left[(\mathbf{K}\mathbf{U}^*)'  \iota_{\mathcal{Y}}\widehat{\boldsymbol{\Omega}}_{\mathbf{Y}} \iota_{\mathcal{Y}}' (\mathbf{K}\mathbf{U}^*) - \iota_{\mathcal{Y}}\diag(\alpha_1, \dots, \alpha_{D-Q_1-1}) \iota_{\mathcal{Y}}' \right] \\
& = (n_y-1) \left[(\mathbf{U}^{\hookrightarrow})'  \iota_{\mathcal{Y}}\widehat{\boldsymbol{\Omega}}_{\mathbf{Y}} \iota_{\mathcal{Y}}' \mathbf{U}^{\hookrightarrow} - \iota_{\mathcal{Y}}\diag(\alpha_1, \dots, \alpha_{D-Q_1-1}) \iota_{\mathcal{Y}}' \right] \\
& = \iota_{\mathcal{Y}} \, (n_y-1) \underbrace{\left[ \mathbf{U} \widehat{\boldsymbol{\Omega}}_{\mathbf{Y}} \mathbf{U}' - \diag(\alpha_1, \dots, \alpha_{D-Q_1-1}) \right]}_{= \mathbf{A}}\, \iota_{\mathcal{Y}}' ; .
\end{split}
\end{equation}
Similarly we define 
\begin{displaymath}
\begin{split}
\widetilde{\mathbf{B}} & := (n_z-1) \left[(\mathbf{K}\mathbf{V}^*)'  \iota_{\mathcal{Z}}\widehat{\boldsymbol{\Omega}}_{\mathbf{Z}} \iota_{\mathcal{Z}}' (\mathbf{K}\mathbf{V}^*) - \iota_{\mathcal{Z}}\diag(\beta_1, \dots, \beta_{D-Q_2-1}) \iota_{\mathcal{Z}}' \right] \\
& = \iota_{\mathcal{Z}}\, (n_z-1) \underbrace{\left[ \mathbf{V} \widehat{\boldsymbol{\Omega}}_{\mathbf{Z}} \mathbf{V}' - \diag(\beta_1, \dots, \beta_{D-Q_2-1}) \right]}_{= \mathbf{B}} \, \iota_{\mathcal{Z}}' ; .
\end{split}
\end{displaymath}
By applying Theorem \ref{theo:approx_eigen_covmat} to the matrices in Eqs.~\eqref{eq:wishart_Y} and ~\eqref{eq:wishart_Z} we obtain that, up to second order terms in the element of $(n_y-1)^{-1}\mathbf{A}$ and $(n_z-1)^{-1}\mathbf{B}$, respectively, it holds
\begin{equation}\label{eq:approx_alpha}
\begin{split}
(n_y -1) \sum_{i=1}^K \widehat{\alpha}_i & \simeq (n_y-1)\sum_{i=1}^K \alpha_i + \sum_{i=1}^K a_{ii}+ \sum_{i=1}^K \sum_{j=K+1}^{D-Q_1-1} \frac{a_{ij}^2}{(n_y-1)(\alpha_i - \alpha_j)}\\
& = (n_y-1) \sum_{i=1}^K \widetilde{\alpha}_i + \sum_{i=1}^K \widetilde{a}_{ii} + \sum_{i=1}^K \sum_{j=K+1}^{D-1} \frac{(\widetilde{a}_{ij})^2}{(n_y-1)(\widetilde{\alpha}_i - \widetilde{\alpha}_j)} 
\end{split}
\end{equation}
and 
\begin{equation}\label{eq:approx_beta}
\begin{split}
(n_z -1)  \sum_{i=1}^K \widehat{\beta}_i & \simeq (n_z-1) \sum_{i=1}^K \beta_i + \sum_{i=1}^K b_{ii} + \sum_{i=1}^K \sum_{j=K+1}^{D-Q_2-1} \frac{b_{ij}^2}{(n_z-1)(\beta_i - \beta_j)} \\
& = (n_z-1) \sum_{i=1}^K \widetilde{\beta}_i + \sum_{i=1}^K \widetilde{b}_{ii} + \sum_{i=1}^K \sum_{j=K+1}^{D-1} \frac{\widetilde{b}_{ij}^2}{(n_z-1)(\widetilde{\beta}_i - \widetilde{\beta}_j)} \; .
\end{split}
\end{equation}
In the previous equations we exploited that from Definition \ref{def:operators_iota} it follows 
\begin{displaymath}
\widetilde{\alpha}_i = \left\{ 
\begin{array}{cl} 
\alpha_{i} & \text{for } i \in \{1, \dots, D-Q_1-1\} \\
0  &  \text{for } i \in \{ D-Q_1, \dots, D-1 \} \; ,  
\end{array} \right.
\end{displaymath}

\begin{displaymath}
\widetilde{\beta}_i = \left\{ 
\begin{array}{cl} 
\beta_{i} & \text{for } i \in \{1, \dots, M-1\} \\
0  &  \text{for } i \in \{ M, \dots, M-1+Q_2 \} \\
\beta_{i-Q_2} &  \text{for } i \in \{  M+Q_2, \dots, D-1 \}
\end{array} \right.
\end{displaymath}

and similarly, for all $i=1, \dots, K$,
\begin{displaymath} 
\widetilde{a}_{ij} = \left\{ 
\begin{array}{cl} 
a_{ij} & \text{for } j \in \{1, \dots, D-Q_1-1 \} \\
0  &  \text{for } j \in \{ D-Q_1, \dots, D-1 \}
\end{array} 
\right. \, , 
\end{displaymath}

\begin{displaymath} 
\widetilde{b}_{ij} = \left\{ 
\begin{array}{cl} 
b_{ij} & \text{for } j \in \{1, \dots, M-1 \}\\
0  &  \text{for } j \in \{M, \dots, M-1+Q_2 \} \\
b_{i(j-Q_2)} & \text{for } j \in \{M+Q_2, \dots, D-1 \}
\end{array} 
\right. \, , 
\end{displaymath}
We further observe that
\begin{displaymath}
\begin{split}
\mathbf{U}^* & \widetilde{\mathbf{A}} \mathbf{U}{^*}' + \mathbf{V}^* \widetilde{\mathbf{B}} \mathbf{V}{^*}'= \\
& = (n_y-1) \mathbf{K}' \; \iota_{\mathcal{Y}} \widehat{\boldsymbol{\Omega}}_{\mathbf{Y}} \iota_{\mathcal{Y}}' \; \mathbf{K} - (n_y-1) \mathbf{U}^* \; \iota_{\mathcal{Y}} \diag(\alpha_1, \dots, \alpha_{D-Q_1-1}) \iota_{\mathcal{Y}}' \; \mathbf{U}{^*}' \\
& \quad + (n_z-1) \mathbf{K}' \; \iota_{\mathcal{Z}} \widehat{\boldsymbol{\Omega}}_{\mathbf{Z}} \iota_{\mathcal{Z}}' \; \mathbf{K} - (n_z-1) \mathbf{V}^* \; \iota_{\mathcal{Z}} \diag(\beta_1, \dots, \beta_{D-Q_2-1}) \iota_{\mathcal{Z}}' \; \mathbf{V}{^*}' \\
& = \mathbf{K}' \left[ (n_y-1) \iota_{\mathcal{Y}} \widehat{\boldsymbol{\Omega}}_{\mathbf{Y}} \iota_{\mathcal{Y}}' + (n_z-1) \iota_{\mathcal{Z}} \widehat{\boldsymbol{\Omega}}_{\mathbf{Z}} \iota_{\mathcal{Z}}' \right] \mathbf{K}\\
& \quad - \mathbf{K}' \left[ (n_y-1) \iota_{\mathcal{Y}} \boldsymbol{\Omega}_{\mathbf{Y}} \iota_{\mathcal{Y}}' + (n_z-1) \iota_{\mathcal{Z}} \boldsymbol{\Omega}_{\mathbf{Z}} \iota_{\mathcal{Z}}' \right] \mathbf{K}\\
& = \mathbf{K}' \left[ (n_y-1) \iota_{\mathcal{Y}} \widehat{\boldsymbol{\Omega}}_{\mathbf{Y}} \iota_{\mathcal{Y}}' + (n_z-1) \iota_{\mathcal{Z}} \widehat{\boldsymbol{\Omega}}_{\mathbf{Z}} \iota_{\mathcal{Z}}' \right] \mathbf{K} - (n_y+n_z-2) \mathbf{K}' \boldsymbol{\Omega}_{pool} \mathbf{K} \\
& = \mathbf{K}' \left[ (n_y-1) \iota_{\mathcal{Y}} \widehat{\boldsymbol{\Omega}}_{\mathbf{Y}} \iota_{\mathcal{Y}}' + (n_z-1) \iota_{\mathcal{Z}} \widehat{\boldsymbol{\Omega}}_{\mathbf{Z}} \iota_{\mathcal{Z}}' \right] \mathbf{K} - (n_y+n_z-2) \diag(\psi_1, \dots, \psi_{D-1}) \, .
\end{split}
\end{displaymath}

Hence, following the same procedure described in \cite{schott1988}, by exploiting the block diagonal form of $\mathbf{U}^*$ and $\mathbf{V}^*$, it is possible to obtain 

\begin{equation}\label{eq:approx_gamma}
\begin{split}
    \sum_{i=1}^K \widehat{\gamma}_i &  = \sum_{i=1}^K \lambda_i\left(\mathbf{K}' \left[ (n_y-1) \iota_{\mathcal{Y}} \widehat{\boldsymbol{\Omega}}_{\mathbf{Y}} \iota_{\mathcal{Y}}' + (n_z-1) \iota_{\mathcal{Z}} \widehat{\boldsymbol{\Omega}}_{\mathbf{Z}} \iota_{\mathcal{Z}}' \right] \mathbf{K}\right) \\
& \simeq (n_y + n_z -2) \sum_{i=1}^K \psi_i + \sum_{i=1}^K (\widetilde{a}_{ii}+\widetilde{b}_{ii})+ \sum_{i=1}^K \sum_{j=K+1}^{D-1} \frac{(\mathbf{u}_i^* \widetilde{\mathbf{A}}_{12}\mathbf{u}_j{^*}'+\mathbf{v}_i^* \widetilde{\mathbf{B}}_{12}\mathbf{v}_j{^*}')^2}{(n_y+n_z-2)(\psi_i-\psi_j)}
\end{split}
\end{equation}
where we denoted with $\mathbf{u}_1^*, \dots, \mathbf{u}_K^*$ and $\mathbf{v}_1^*, \dots, \mathbf{v}_K^* \in \mathbb{R}^K$ the rows of $\mathbf{U}^*_1$ and $\mathbf{V}^*_1$, with $\mathbf{u}_{K+1}^*, \dots, \mathbf{u}_{D-1}^*$ and $\mathbf{v}_{K+1}^*, \dots, \mathbf{v}_{D-1}^* \in \mathbb{R}^{D-K-1}$ the rows of $\mathbf{U}^*_2$ and $\mathbf{V}^*_2$, and with $\widetilde{\mathbf{A}}_{12}$ and $\widetilde{\mathbf{B}}_{12}$ the matrices of size $K \times (D-K-1)$ comprising the first $K$ rows and the last $D-K-1$ columns of $\widetilde{\mathbf{A}}$ and $\widetilde{\mathbf{B}}$, respectively. 

By replacing the approximation derived in Eqs.~\eqref{eq:approx_alpha}, \eqref{eq:approx_beta}, and \eqref{eq:approx_gamma} into the definition of the test statistic $T$ provided in Eq.~\eqref{eq:def_T} and by exploiting Eq.~\eqref{eq:rel_latentval_H0} we obtain that under $H_0$ we can approximate
\begin{equation}\label{eq:approx_T}
    T \simeq \sum_{i=1}^K \sum_{j=K+1}^{D-1} \left[ \frac{(\widetilde{a}_{ij})^2}{(n_y-1)(\widetilde{\alpha}_i - \widetilde{\alpha}_j)} + \frac{\widetilde{b}_{ij}^2}{(n_z-1)(\widetilde{\beta}_i - \widetilde{\beta}_j)} - \frac{(\mathbf{u}_i^* \widetilde{\mathbf{A}}_{12}\mathbf{u}_j{^*}'+\mathbf{v}_i^* \widetilde{\mathbf{B}}_{12}\mathbf{v}_j{^*}')^2}{(n_y+n_z-2)(\psi_i-\psi_j)} \right] \, .
\end{equation}
We further assume $\boldsymbol{\Omega}_{\mathbf{Y}}$ and $\boldsymbol{\Omega}_{\mathbf{Z}}$ to be diagonal and we define, for all $i=1, \dots, D-1$, $\widetilde{\psi}_i = \frac{(n_y-1) \widetilde{\alpha}_i + (n_z-1) \widetilde{\beta}_i}{n_y+n_z-1}$, which are an unordered version of the eigenvalues $\{\psi_i \}_{i=1 \dots, D-1}$ of $\boldsymbol{\Omega}_{pool}$. A permutation $\sigma: \{1, \dots, D-1 \} \longrightarrow \{1, \dots, D-1 \}$ exists such that $\psi_i = \widetilde{\psi}_{\sigma(i)}$, $\sigma(i) = i$ for all $i=1, \dots, M-1$, and  $\mathbf{K} = \mathbf{U}^* = \mathbf{V}^* = \boldsymbol{\Pi}_{\sigma}$, being $(\boldsymbol{\Pi}_{\sigma})_{ij} = \delta_{j\sigma(i)}$.
Hence the approximation of $T$ becomes 
\begin{equation}\label{eq:approx_T_diagonalcase}
T \simeq \sum_{i=1}^K \sum_{j=K+1}^{D-1} \frac{(n_y-1)(n_z-1)\left( \widetilde{a}_{ij} (\widetilde{\beta}_i - \widetilde{\beta}_j)/(n_y-1) - \widetilde{b}_{ij} (\widetilde{\alpha}_i - \widetilde{\alpha}_j)/(n_z-1)\right)^2}{(\widetilde{\alpha}_i - \widetilde{\alpha}_j)(\widetilde{\beta}_i - \widetilde{\beta}_j)[(n_y-1)(\widetilde{\alpha}_i - \widetilde{\alpha}_j) + (n_z-1)(\widetilde{\beta}_i - \widetilde{\beta}_j)]} \, .
\end{equation}
It can be easily shown that the right-hand side of Eq.~\eqref{eq:approx_T_diagonalcase} is asymptotically distributed as a linear combination of independent chi-squared random variables. Therefore, as suggested by \cite{schott1988}, also in the presence of structural zeros the test statistics $T$ can be approximated as in Eq.~\eqref{eq:final_approx_T} where $\mu_T$ and $\sigma^2_T$ are obtained be computing mean and variance of the right-hand side of Eq.~\eqref{eq:approx_T}. 
\end{proof}


 \bibliographystyle{elsarticle-harv} 
 \bibliography{Bibliografia}








\end{document}